\documentclass[pra,twocolumn,letterpaper,showpacs,superscriptaddress,amsmath]{revtex4}
\usepackage{graphicx}
\usepackage{subfigure} 
\usepackage{dcolumn}
\usepackage{bm}

\newcommand{\ket}[1]{\mbox{$|#1\rangle$}}
\newcommand{\bra}[1]{\mbox{$\langle#1|$}}
\newcommand{\braket}[3]{\Big\langle#1 \Big| #2 \Big| #3\Big\rangle}

\newcommand{\lb}{\left[}
\newcommand{\rb}{\right]}
\newcommand{\lp}{\left(}
\newcommand{\rp}{\right)}

\newcommand{\pfun}{\mathcal{Z}}
\newcommand{\rhoth}{\rho_\text{th}}
\newcommand{\rhoeff}{\rho_\text{eff}}
\newcommand{\rhoch}{\rho_\text{cH}}
\newcommand{\rhocf}{\rho_\text{cf}}
\newcommand{\rhob}{\rho_\text{b}}
\newcommand{\ubell}{U_\text{b}}
\newcommand{\uch}{U_\text{cH}}
\newcommand{\ucnot}{U_\text{cnot}}
\newcommand{\lmin}{\lambda_\text{min}}
\newcommand{\alphatr}{\alpha_\text{tr}}
\newcommand{\vmin}{v_\text{min}}
\newcommand{\ufan}{U_\text{fan}}
\newcommand{\eunit}{\mathcal{E}_\ell}

\newcommand{\negm}{E_\mathcal{N}}
\newcommand{\jmax}{j_\text{max}}

\begin{document}

\title{Bounds on the entanglability of thermal states\\ in liquid-state nuclear magnetic resonance} 

\author{Terri M.~Yu} 
\altaffiliation{Current address: Department of Physics, Yale University, New Haven, CT 06520.}
\affiliation{Center for Bits and Atoms - MIT Media Laboratory, Cambridge, Massachusetts 02139}
\affiliation{Department of Electrical Engineering and Computer Science, Massachusetts Institute of Technology, Cambridge, Massachusetts 02139}
\author{Kenneth R.~Brown}
\affiliation{Center for Bits and Atoms - MIT Media Laboratory, Cambridge, Massachusetts 02139}
\affiliation{Department of Chemistry, University of California, Berkeley, California 94720} 
\author{Isaac L.~Chuang}
\affiliation{Center for Bits and Atoms - MIT Media Laboratory, Cambridge, Massachusetts 02139}
\affiliation{Department of Physics, Massachusetts Institute of Technology, Cambridge, Massachusetts 02139}


\begin{abstract}
The role of mixed state entanglement in liquid-state nuclear magnetic resonance (NMR) quantum computation is not yet well-understood.  In particular, despite the success of quantum information processing with NMR, recent work has shown that quantum states used in most of those experiments were not entangled.  This is because these states, derived by unitary transforms from the thermal equilibrium state, were too close to the maximally mixed state.  We are thus motivated to determine whether a given NMR state is entanglable -- that is, does there exist a unitary transform that entangles the state?  The boundary between entanglable and nonentanglable thermal states is a function of the spin system size $N$ and its temperature $T$.  We provide new bounds on the location of this boundary using analytical and numerical methods; our tightest bound scales as $N\sim T$, giving a lower bound requiring at least $N\sim 22,000$ proton spins to realize an entanglable thermal state at typical laboratory NMR magnetic fields.  These bounds are tighter than known bounds on the entanglability of effective pure states.
\end{abstract}

\maketitle


\section{Introduction}

Entanglement is hidden information in the form of nonlocal correlations between two or more quantum degrees of freedom.  Recent work supports the view that entanglement is a quantifiable physical resource.  Entangled states are useful for communicating quantum information~\cite{Bennett93a, Bennett96a, Bennett96b}.  It also appears that entangled states are required to exploit the exponential speedup of many pure-state quantum algorithms~\cite{Azuma01a,Braunstein02a}, including Shor's algorithm~\cite{Ekert98a, Linden01a, Jozsa03a}.

Quantum computers based on liquid-state nuclear magnetic resonance (NMR) have successfully implemented these same quantum algorithms~\cite{Jones98a, Jones98b, Chuang98a, Chuang98b, Vandersypen01a}.  However, Braunstein \textit{et al.}~have shown that the experiments use separable states~\cite{Braunstein99a}.  The initial NMR states are not pure, but rather mixed states derived from a thermal ensemble.  Due to the small number of qubits and the weak initial polarization of the ensemble, these effective pure states are too random to possess any entanglement.

This startling result raises doubts about the validity of NMR quantum computation.  How can NMR techniques demonstrate quantum algorithms without entanglement?  Are entangled states necessary resources for quantum computation?  These questions are yet to be conclusively answered, although it has been pointed out that 1) the essential role of entanglement in pure-state algorithms does not necessarily extend to the mixed states used in NMR~\cite{Laflamme02a, Jozsa03a} and that 2) there exist efficient pure-state quantum algorithms without entanglement~\cite{Meyer00a}.  In fact, a model of computation using one qubit purified from a mixed state has been shown to be capable of performing non-classical algorithms~\cite{Knill98a}.  Furthermore, attempts to construct classical models of NMR quantum computation have thus far suffered from exponential scaling deficiencies~\cite{Schack99a, Menicucci02a}.  Some speculate that entangling operations --- rather than entangled states --- are the source of exponential speedup in quantum computers~\cite{Nielsen98a}.  

Anwar \textit{et al.}~recently implemented the Deutsch-Jozsa and Grover algorithms on a nearly pure two-qubit NMR state~\cite{Anwar04b, Anwar04c}, proving that it is possible to perform NMR quantum computation with entanglement.  These results are impressive, but it will be challenging to scale the system since each qubit must be selectively excited.  Moreover, as the authors state~\cite{Anwar04a}, the experiment cannot answer fundamental questions regarding the role of mixed state entanglement in quantum computation.

We attempt to address these questions through a theoretical study of entanglement in thermal states, the natural mixed states in liquid-state NMR systems.  Due to reasons we explain, the thermal state might be more easily entangled than the effective pure state used in current NMR quantum computation experiments~\footnote{Throughout this paper, we use the term ``NMR'' to mean ``liquid-state NMR'' unless explicitly noted.}.  

The spins in a liquid-state NMR system are largely uncoupled, although the tiny coupling allows a universal set of unitary operations to be performed on the spins.  We thus approximate the initial state of the NMR system with a fully uncoupled Hamiltonian.  For ease in our analysis, we also assume that all the spins are identical.  The thermal state corresponding to such a system is then completely specified by the number of qubits (spins) $N$ and the dimensionless quantity $\alpha$, which is a measure of the state's polarization.  The polarization is the bias of the spins towards parallel alignment with the magnetic field; it is also approximately invesely proportional to the equilibrium temperature $T$.  Finally, we assume that an arbitrary unitary transform can be efficiently performed on each $N$ qubit system in the ensemble. 

Here we present bounds on the boundary in $N$-$\alpha$ parameter space between thermal states that can and cannot be entangled.  A lower/upper bound on the boundary demarcates parameter space where thermal states can/cannot be entangled.  We obtain a lower bound on thermal states that is tighter than previous calculations for effective pure states~\cite{Braunstein99a, Dur00a}.  From a practical standpoint, a lower bound on the parameter space boundary is equivalent to an upper bound on the minimum number of qubits and polarization required to create an entangled NMR state in the laboratory.

We begin in Sec.~\ref{sec:review} with a brief overview of previous work on NMR state entanglement.  Then we assess different methods for entangling NMR states in Sec.~\ref{sec:exp} and describe our approach to calculating and bounding the entanglement of transformed thermal states in Sec.~\ref{sec:approach}.  Next, we present the results of this paper.  In Sec.~\ref{sec:npt}, we use the negative partial transpose test~\cite{Peres96a, Horodecki96f} to calculate bounds on the entanglement of thermal states transformed by a specific Bell unitary.  In Sec.~\ref{sec:bounds}, we develop a general method to calculate bounds on the entanglement of generic Bell-transformed thermal states, based on a classification scheme by D\"ur and Cirac~\cite{Dur00a}.  Combining the method with majorization theory could potentially lead to tighter bounds.  This possibility is discussed in Sec.~\ref{sec:majorization}.  Finally, we summarize our results and conclude in Sec.~\ref{sec:conclusion}.


\section{Review of bounds on NMR state entanglement} \label{sec:review}

We first briefly review mixed state entanglement.  A mixed state $\rho$ shared by party $A$ and party $B$ is separable if and only if it can be expressed as
\begin{equation}
\rho = \sum_i p_i~\rho_i^A \otimes\rho_i^B \label{eq:mixedent2-rho} 
\end{equation}
where $\rho_i^A$ and $\rho_i^B$ are density matrices contained in the Hilbert spaces of parties $A$ and $B$ respectively and the weights $p_i$ are probabilities.  The bipartite state $\rho$ is entangled if it cannot be written in this form.

In general, a mixed state $\rho$ shared by $K$ parties is separable if and only if it can be expressed as
\begin{equation}
\rho = \sum_i p_i \bigotimes_{j=1}^K \rho_i^j \label{eq:mixedentN}
\,.
\end{equation}
Each density matrix $\rho_i^j$ is in the Hilbert space of the $j$th party.  The $K$-partite state $\rho$ is entangled if it cannot be written in the form of Eq.~(\ref{eq:mixedentN}).  If $K$ is equal to the number of degrees of freedom (e.g. the number of qubits) and $\rho$ satisfies Eq.~(\ref{eq:mixedentN}), then we call $\rho$ fully separable.

We desire a way to determine whether a state is entangled.  The above definitions for entanglement are a possible starting point, but they require us to examine an infinite number of decompositions.   For bipartite mixed states, there exists a simple, computable criterion for entanglement~\cite{Peres96a, Horodecki96f}:
\begin{quote}
If the partial transpose of a bipartite state $\rho$ has a negative
eigenvalue, $\rho$ is entangled.
\end{quote}
Note that this criterion gives only a necessary condition for entanglement except in the case of two qubits, where it becomes both necessary and sufficient.

The partial transpose operation is defined as follows.  Any bipartite density matrix $\rho$ may be expressed as
\begin{equation}
        \rho = \sum_{a,a^\prime,b,b^\prime} C_{a,a^\prime,b,b^\prime} \ket{a}\bra{a^\prime} \otimes \ket{b}\bra{b^\prime}
\,.
\end{equation}
The basis states $\ket{a}$, $\ket{a^\prime}$ and $\ket{b}$, $\ket{b^\prime}$ are in the Hilbert spaces of parties $A$ and $B$ respectively, and the numbers $C_{a,a^\prime,b,b^\prime}$ are complex and constrained to give $\rho$ unit trace.  The partial transpose of $\rho$ with respect to party $A$ is denoted $\rho^{T_A}$ and given by
\begin{equation}
        \rho^{T_A} = \sum_{a,a^\prime,b,b^\prime} C_{a,a^\prime,b,b^\prime} \ket{a^\prime}\bra{a} \otimes \ket{b}\bra{b^\prime} \label{eq:ptranspose}
\,.
\end{equation}
If $\rho^{T_A}$ has at least one negative eigenvalue, we say that $\rho$ has negative partial transpose (NPT).  Otherwise, we say that $\rho$ has positive partial transpose (PPT).

We now turn to the specific case of entanglement in NMR quantum computation.  Liquid-state NMR quantum computing experiments are performed on samples containing an ensemble of $10^{18}-10^{19}$ identical molecules under a strong, static magnetic field ($\sim$12 T).  Each molecule possesses $N$ spin-1/2 nuclei, which represent the qubits.  The large number of molecules is needed because the signal from a single nuclear spin is exceedingly weak.  The dynamics of the NMR system is dominated by the Zeeman interaction between the nuclear spins and the magnetic field.

The natural quantum state in NMR is the thermal state
\begin{equation}
\rhoth = \frac{e^{-\mathcal{H}/kT}}{\pfun}
\label{eq:thstate} 
\end{equation}
where $\mathcal{H}$ is the Hamiltonian, $k$ is the Boltzmann constant, $T$ is the absolute temperature, and $\pfun$ is the partition function.

The thermal state is highly mixed in liquid-state NMR systems.  Yet most quantum algorithms require pure states.  To circumvent this problem, experimenters first create initial spin states of form
\begin{equation}
\rhoeff = (1-\epsilon) M_d + \epsilon\ket{0}\bra{0} \label{eq:effpstate} 
\,.
\end{equation}
Here the dimension of $\rhoeff$ is $d = 2^N$, $M_d = I_d/d$ is the maximally mixed state ($I_d$ is the $d$-dimensional identity matrix), and $\epsilon$ characterizes the fraction of ground state $\ket{0}$.  These states are called effective pure states or pseudopure states~\cite{Gershenfeld97a, Cory97a} because 1) a unitary transform only evolves the excess ground state population and 2) the term proportional to $M_d$ is unobservable in conventional NMR experiments.

Braunstein \textit{et al.}~established that the states used in all NMR quantum computing experiments up to now have never been entangled~\cite{Braunstein99a}.  They studied near maximally mixed states of the form
\begin{equation}
\rho_\epsilon = (1-\epsilon)M_d + \epsilon\rho^\prime \label{eq:maxentstate}
\end{equation}
where $\rho^\prime$ is an arbitrary density matrix and $\epsilon$ characterizes the fraction of the mixture in $\rho^\prime$~\footnote{We follow the convention of the literature and use the same symbol $\epsilon$ to denote the bias away from $M_d$ in both Eqs.~(\ref{eq:effpstate}) and~(\ref{eq:maxentstate}).}.  Effective pure states clearly fall into this class of states.

The authors established bounds on the entanglement of $\rho_\epsilon$ for a system of $N$ spin-1/2 particles:
\begin{eqnarray}
\epsilon &\leq& \frac{1}{1 + 2^{2N-1}} \Rightarrow \rho_\epsilon \text{\, is always separable} \label{eq:braunsep} \\
\epsilon &>& \frac{1}{1 + 2^{N/2}} \Rightarrow \rho_\epsilon \text{\, can be nonseparable} \label{eq:braunent} 
\,.
\end{eqnarray}
The first expression may be interpreted as a lower bound on the size of the separable neighborhood around $M_d$.  When $\epsilon$ is sufficiently small, $\rho_\epsilon$ is separable.  The second expression may be interpreted as an upper bound on the separable neighborhood.  When $\epsilon$ is sufficiently large, there exists some $\rho^\prime$ such that $\rho_\epsilon$ is nonseparable.

These bounds are applicable to liquid-state NMR because quantum computations can only transform a state of form $\rho_\epsilon$ to another state of form $\rho_\epsilon$.  Liquid-state NMR systems are approximately closed and thus evolve unitarily.  Assuming the largest practical initial polarization, $\epsilon$ may be as large as $3 \times 10^{-5}$ for $N$=2.  Since $\epsilon$ scales as $N/2^N$ for low polarization, Eq.~(\ref{eq:braunsep}) shows that the states used in NMR quantum computation are separable when $N \leq 12$.  Therefore, current NMR experiments, for which $N \leq 7$, do not use entangled states.

Gurvits and Barnum~\cite{Gurvits04a} recently tightened the lower bound to
\begin{equation}
\epsilon \leq \frac{3}{2(6^{N/2})}  \label{eq:gursep}
\,.
\end{equation}
This result shows that the states used in NMR quantum computation are separable when $N < 32$.

There is also a tighter upper bound due to D\"ur and Cirac, but we defer discussion of this bound to Sec.~\ref{sec:bounds}.
 
In what follows, we will be interested in NMR parameter space described by $N$ and $\alpha$ (defined in Sec.~\ref{sec:exp}).  We thus re-express the above bounds in terms of these parameters.

It can be shown~\cite{Knill98a} that $\epsilon$ is given by
\begin{equation}
\epsilon = \frac{e^{N\alpha}}{\pfun} - \frac{1-e^{N\alpha}/\pfun}{2^N-1}
\approx \frac{N\alpha}{2^N-1}
\label{eq:eps-rhoeff}
\end{equation}
with the approximation being valid in the limit $\alpha \ll 1$.  

Using Eq.~(\ref{eq:eps-rhoeff}), the Braunstein \textit{et al.}~bound on entanglable $\rhoeff$ is
\begin{equation}
\alpha > -\frac{1}{2}\ln (\sqrt{2}-1)
\,,
\label{eq:braunent2} 
\end{equation}
and the Gurvits-Barnum bound on nonentanglable $\rhoeff$ is
\begin{equation}
\alpha \leq \frac{3(2^N-1)}{2N(6^{N/2})} 
\,.
\label{eq:gursep2}  
\end{equation}
In this paper, we call a state \textit{entanglable} if there exists a unitary that transforms the state into an entangled one.  This terminology allows us to discuss entanglement without referring to the specific unitary that entangled the original state.


\section{How to entangle an NMR state?} \label{sec:exp}

We wish to address the question of whether it is possible to entangle an NMR thermal state.  In this section, we evaluate several methods for achieving entanglement.  The discussion here motivates our approach in Sec.~\ref{sec:approach}. 

\subsection{Initial NMR state}

The first consideration is the choice of initial state.  Current NMR quantum computation begins with effective pure states, but we simply want to obtain an entangled state by any means.  The thermal state is a better initial state for reasons we now explain.

One possible decomposition of an effective pure state is the following mixture of transformed thermal states:
\begin{equation}
\rhoeff = \sum_{j=1}^{d-1} p_j U_j \rhoth U_j^\dagger \label{eq:rhoeff3} 
\,.
\end{equation}
Here $d$ is the dimension of $\rhoeff$, $p_j = 1/(d-1)$, and $U_j$ are cyclic permutation matrices.

Since the numbers $p_j$ are probabilities, $\rhoeff$ satisfies the mathematical definition of a convex combination.  Now entanglement is generally a convex function and therefore is reduced under convex combination.  Hence, an individual transformed thermal state $U_j \rhoth U_j^\dagger$ may possess more entanglement than an effective pure state of the same dimension.  

These considerations motivate us to focus on entangling NMR \textit{thermal states} in this paper.  

For convenience in the analysis, we assume that all spins in the NMR molecules have the same Zeeman energy splitting and consider an isotropic Hamiltonian 
\begin{equation}
\mathcal{H} = \frac{h\nu}{2}~\sum_{i=1}^N Z_i \label{eq:nmrham}
\,,
\end{equation}
expressed in the computational (spin) basis.  Here $\nu = \gamma B$ is the frequency corresponding to a spin with gyromagnetic ratio $\gamma$ in a magnetic field of strength $B$, $Z_i$ is the Pauli $Z$ operator acting on the $i$th spin, and $N$ is the number of spin-1/2 particles in each molecule.  We neglect the $J$-coupling interaction between neighboring spins as it is $10^{-6}$ times smaller than the Zeeman energy.

The corresponding thermal state is given by Eq.~(\ref{eq:thstate}) and the Hamiltonian above.  Notice that $\rhoth$ is diagonal in the computational basis, invariant under the exchange of any two spins, and fully separable. 

We define $\alpha$, a measure of the thermal state's polarization, to be
\begin{equation}
\alpha \equiv \frac{h\nu}{2kT}
\,.
\label{eq:alpha}
\end{equation}
Notice that when $\alpha \ll 1$, this quantity is approximately the difference between the fraction of spins aligned and the fraction of spins anti-aligned with the magnetic field, i.e. the polarization.  Since NMR quantum computing experiments are performed in the $\alpha \ll 1$ regime, we henceforth refer to $\alpha$ as simply the polarization in this paper. 

Putting together Eqs.~(\ref{eq:thstate}),~(\ref{eq:nmrham}), and~(\ref{eq:alpha}), the matrix elements of the thermal state in the computational basis are
\begin{equation}
\bra{i}\rhoth\ket{j} = \frac{\delta_{ij}}{\pfun} e^{[N-2w(i)]\alpha} 
\label{eq:thstate-diag}
\end{equation}
where the Hamming weight $w(i)$ is the number of 1s in the binary expression for $i$ and $0 \leq i,j \leq 2^N-1$.  In the limit $\alpha \rightarrow 0$, we see that $\rhoth \rightarrow I$.

\subsection{Methods to facilitate entanglement}

An experiment to entangle an NMR thermal state consists of two parts:
\begin{enumerate}
\item Prepare the initial thermal state $\rhoth$ with number of qubits $N$ and polarization $\alpha$.
\item Apply a unitary operation to entangle the thermal state: $\rhoth \mapsto U\rhoth U^\dagger$.
\end{enumerate}

In view of the above procedure, there are several ways we can facilitate the creation of entangled thermal states:
\begin{itemize}
\item Enhance initial polarization 
\item Increase the number of qubits (spins) per molecule
\item Perform algorithmic cooling 
\item Find optimally entangling unitary operations.
\end{itemize}

We expect that increased polarization will yield a thermal state that is more easily entangled.  Doing so moves the thermal state away from the maximally mixed state.  Increasing the number of qubits should have the same effect, but the intuition behind this conjecture is weaker.  As we have seen, the bounds on the nonentanglability of $\rhoeff$ relax as $N$ increases.  Moreover, the volume of separable states appears to decrease with Hilbert space dimension.   A lower bound for the volume of separable states~\cite{Vidal99a} has been found, which decreases exponentially with $N$.  Numerical evidence~\cite{Zyczkowski98a} supports the same trend.

We now examine the viability of each method.  

\subsubsection{Enhancing initial polarization}

We discuss three experimental approaches for increasing the initial polarization: 1) lowering the temperature, 2) increasing magnetic field strength, and 3) chemical and optical methods.  As a reference, a state-of-the-art liquid-state NMR experiment using seven proton nuclear spin molecules at room temperature has parameters $N = 7$, $\alpha = 4 \times 10^{-5}$, $B = 12$ T, and $T = 300$ K.  A seven-qubit effective pure state must have $\alpha > 3 \times 10^{-2}$ to be outside of the nonentanglable regime described by Eq.~(\ref{eq:gursep2}).

Lowering the temperature of the experiment is impractical because any appreciable improvement in polarization demands such a dramatic drop in temperature that the sample becomes solid.  Unless the spin system is dilute, it acquires large dipolar couplings that smear out the spectrum and make selective excitation of spins difficult.  A dilute spin system could be used, but we ideally desire large $N$.  There are several proposals for solid-state NMR quantum computers that address these problems~\cite{Yamaguchi99a, Cory00a}, but none have been experimentally realized thus far. 

Using a stronger magnetic field as a way to increase polarization is also limited.  Because present NMR magnets are based on superconducting coils that are current-limited, the largest attainable field is 21 T.  Higher magnetic fields can be achieved with DC Bitter resistive magnets (33-45 T) and capacitively-driven magnets (50-60 T), but the generated fields are too spatially inhomogeneous to be used in liquid-state NMR.  The magnets also consume massive power and require extensive cooling.

Inducing higher polarization by chemical and optical means is a much more promising approach and has already been experimentally demonstrated.   Optically pumped xenon has been shown to give a tenfold increase in $\alpha$ for a two qubit molecule~\cite{Verhulst01a}.  Another factor of three or four may be gained if the xenon is isotopically pure.  Further increases in $\alpha$ are also possible if higher polarizations of xenon are used.  The experiment of Ref.~\cite{Verhulst01a} used 1\% polarized xenon, but Xe polarizations as high as 67\% have been obtained~\cite{Zook02a}.  Methods involving \textit{para} hydrogen (two protons in the singlet state) yield even larger polarization.  By reacting \textit{para} hydrogen with another molecule, Anwar \textit{et al.}~\cite{Anwar04a} have recently created two-qubit effective pure states with $\epsilon = 0.916 \pm 0.019$.  These initial states are entangled according to Eq.~(\ref{eq:braunent}).

\subsubsection{Increasing the number of qubits} 

Entanglement can also be achieved by using more qubits, but this approach is limited by substantial fundamental problems~\cite{Jones00a}.  Since individual spins are addressed by selective excitation, the resonance frequency of each spin must be well-separated from the others.  This requirement may be satisfied by using nuclear spins of different chemical species.  There are five distinct spin-1/2 nuclei commonly used for liquid-state NMR: $^1$H, $^{13}$C, $^{15}$N, $^{19}$F, and $^{31}$P.  Spins of the same type may also be used, but the largest number of homogeneous spins demonstrated in an experiment thus far~\cite{Linden99a} is $N=6$.  Thus, for conventional liquid-state NMR quantum computation, the requirement of selective excitation restricts experiments to $N=30$.  It may be possible to improve upon this limit by using polymers~\cite{Lloyd93a} or higher-order spins.  Classical logic operations have been experimentally demonstrated in spin-3/2~\cite{Sinha01a} and spin-7/2 nuclei~\cite{Murali02a}. 

\subsubsection{Algorithmic cooling}

Algorithmic cooling is a compression technique that concentrates the polarization of many qubits into a smaller number of qubits.  A well-known example is the Schulman-Vazirani procedure~\cite{Schulman99a}, which starts with $N_0$ qubits in thermal equilibrium at polarization $\alpha_0$ and extracts from these initial qubits a string of order $(\alpha_0)^2 N_0$ qubits each with $p_0 > 1 - \frac{1}{2}N_0^{-10}$ where $p_0$ is the probability of measuring $\ket{0}$.  This method produces very pure qubits but at a high cost for the weak polarizations in current NMR experiments.  For $\alpha_0=4 \times 10^{-5}$, we need approximately $6.3\times 10^{8}$ initial thermal qubits to obtain just one pure qubit.  The enormous number of initial qubits is needed because the Schulman-Vazirani procedure preserves the entropy of the system.  If entropy conservation is circumvented, better results are possible.  Another scheme proposes to use an external set of spins that act as a heat bath~\cite{Boykin02a}.  This method can create effective pure states with initial to final qubit ratio of 50.   

If $\alpha$ can be boosted with some of the techniques previously mentioned, the Schulman-Vazirani procedure becomes more realistic.  For example, beginning with $N_0 = 30$ and $\alpha_0 = 0.2$, we can obtain one nearly pure qubit.

\subsubsection{Entangling unitary operations}

Entangling NMR states via unitary operations has been largely unexplored.  The thermal state is separable, but we can apply a unitary transform to move $\rhoth$ into an entangled region in Hilbert space.  It is currently unknown what unitaries optimally entangle mixed states except for the case of two qubits~\cite{Verstraete01a}.  One group has proposed quantitative measures for the power of entangling operators~\cite{Nielsen03a}.    


\section{Approach} \label{sec:approach}

We choose to investigate the approach of applying a unitary transform to entangle the thermal state.  More specifically, we are interested in the question of whether a given $N$-qubit thermal state with initial polarization $\alpha$ is entanglable, that is, does there exist a unitary that entangles the state?

Studying this problem will be fruitful for two reasons.  First, we can address fundamental questions regarding NMR state entanglement by examining the scaling behavior of entanglable regions in $N$-$\alpha$ parameter space.  Second, our study provides insight as to how experiments may be designed to realize entangled NMR states.

To establish whether a thermal state is entanglable, we must in principle search over the space of all unitary operations.  Rather than attempt this intractable procedure, we focus our attention on the family of transformations $\ubell$ that maps the computational basis to the Bell state basis.  We choose $\ubell$ as a starting point because it generates maximal entanglement when applied to the ground state.  The thermal state has the greatest population in the ground state, so $\ubell$ may be effective in entangling $\rhoth$.  As we shall see, the Bell unitaries also possess symmetries that make them particularly amenable to analysis.   

\begin{figure}[t]
\includegraphics[width=0.45\textwidth]{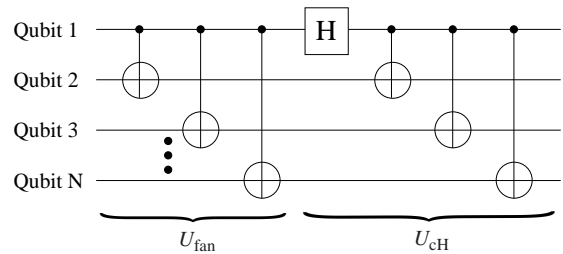}
\caption[short list entry]{Quantum circuit for $\uch\ufan$ unitary operation.  The $\ufan$ unitary acts like a fanout gate on computational states.  The $\uch$ unitary transforms computational states into Bell states.}
\label{fig:uchufan}
\end{figure}

We first define a ``standard'' Bell unitary: the controlled NOT-Hadamard transform $\uch$ depicted in the right half of Fig.~\ref{fig:uchufan}.  A Hadamard gate is applied to the first qubit followed by a set of controlled-NOT (CNOT) gates on the rest of the qubits.  The other unitaries in the Bell transformation family are permutations of $\uch$:
\begin{equation}
\ubell(P) = P \uch P^\dagger
\end{equation}
with $P$ being a matrix that permutes the computational state to Bell state mapping.  

We will later study another Bell unitary $\uch\ufan$, shown in Fig.~\ref{fig:uchufan}.  The permutation matrix $\ufan$ is also known as the fanout gate because it bit flips the latter $N-1$ qubits if the first qubit is in state $\ket{1}$. 

For notational convenience, we also define the following transformed states:
\begin{eqnarray}
\rhob &\equiv& \ubell\rhoth\ubell^\dagger \\
\rhoch &\equiv& \uch\rhoth\uch^\dagger \\
\rhocf &\equiv& \uch\ufan\rhoth\ufan^\dagger\uch^\dagger 
\,.
\end{eqnarray}

Our task is now to determine if a given Bell-transformed thermal state $\rhob$ is entangled.  One method is to apply the negative partial transpose test.  This approach is studied in Sec.~\ref{sec:npt}.  

If all the eigenvalues of partial transposed state can be calculated, we may quantify the amount of entanglement by the negativity entanglement measure~\cite{Zyczkowski98a, Vidal02a}.  The negativity of a state $\rho$ is defined as
\begin{equation}
\negm(\rho) = \frac{\|\rho^{T_A}\|_1 - 1}{2}
\,.
\end{equation}
Here $\|\rho^{T_A}\|_1$ denotes the trace norm of $\rho^{T_A}$, which is defined as $\text{tr}(\rho^{T_A\dagger}\rho^{T_A})$.  For diagonalizable operators, the trace norm reduces to the sum of the absolute eigenvalues $\sum_i |\lambda_i|$.  Negativity has the range $0 \leq \negm(\rho) \leq 1/2$, taking on the smallest value for separable $\rho$ and the largest value for maximally entangled pure states.  Compared to other known entanglement measures~\cite{Bennett96a, Vedral98a}, it has the advantage of being relatively simple to compute for arbitrarily large $N$.  

The partial transpose test may give different results depending on how the qubits are distributed between parties $A$ and $B$.  The thermal state is totally symmetric, and the unitary transforms we examine in this paper are symmetric among the latter $N-1$ spins.  Therefore, it is sufficient to consider partitions where the first $q$ spins are assigned to party $A$ and the remaining $N-q$ spins to party $B$.  We write such a bipartite split as $\{q, N-q\}$.  

A more general method for determining the entanglability of thermal states is the following.  Suppose we have a transformed thermal state whose entanglement is difficult to determine, but we also have a state $\rho^\prime$ whose entanglement is easy to classify.  If we can find a quantum operation that converts the transformed thermal state into $\rho^\prime$ without generating any new entanglement, then the entanglement of $\rho^\prime$ bounds the entanglement of the transformed thermal state.  This approach is studied in Sec.~\ref{sec:bounds}.  


\section{NPT bounds on the entanglement of CNOT-Hadamard-transformed thermal states} \label{sec:npt}

We first derive a formula for the negativity of the CNOT-Hadamard-transformed thermal states under the $\{1,N-1\}$ bipartite split.  Then we find an NPT bound on entangled CNOT-Hadamard-transformed thermal states under the $\{N/2,N/2\}$ bipartite split.  This bound is empirically derived, but it will be confirmed by the analysis of Sec.~\ref{sec:bounds}.

\subsection{Negativity formula for $\{1, N-1\}$ bipartite split} 

The quantum circuit for $\uch$ (see Fig.~\ref{fig:uchufan}) suggests that the first qubit is special.  The thermal state may be rewritten to separate the behavior of the first qubit from the others:
\begin{eqnarray}
\rhoth &=& \frac{1}{\pfun}~\exp\lb -\lp \sum_{i=1}^N Z_i \rp\alpha\rb \nonumber \\
       &=& \frac{1}{\pfun}~\exp\big(-Z_1\alpha\big)\otimes\exp\lb -\lp \sum_{i=2}^N Z_i \rp\alpha\rb 
\label{eq:thermal1}
\end{eqnarray}
where we have used Eqs.~(\ref{eq:thstate}),~(\ref{eq:nmrham}), and~(\ref{eq:alpha}).  Operators with a subscript 1 are taken to be acting on the first qubit. 

We then insert a complete set of angular momentum states $\ket{j,m}\bra{j,m}$ to represent the collective spin state of the latter $N-1$ qubits.  The symbol $j$ denotes the total spin angular momentum whereas $m = -j,-j+1, \ldots, j-1, j$ denotes the total azimuthal spin angular momentum.  As a result, the above equation becomes
\begin{equation}
\rhoth = \frac{1}{\pfun} e^{Z_1\alpha} \otimes \sum_{j,m} e^{-2m\alpha} \ket{j,m}\bra{j,m}
\,.
\end{equation}
Because $\lp\sum_{i=2}^N Z_i/2\rp \ket{j,m} = m \ket{j,m}$, each angular momentum term is multiplied by $e^{-2m\alpha}$.  

Now we examine the effect of $\uch$ on the thermal state.  It is easier to analyze the problem if we break the unitary down into the application of a Hadamard gate on the first qubit ($H_1$) followed by a collective CNOT on the other $N-1$ qubits ($\ucnot$), i.e.~$\uch = \ucnot H_1$. 

Applying a Hadamard operation to the first qubit of $\rhoth$ yields the state
\begin{eqnarray}
\rhoth^\prime &=& H_1 \rhoth H_1^\dagger \nonumber \\
&=& \frac{1}{\pfun} e^{X_1\alpha}\otimes\sum_{j,m} e^{-2m\alpha} \ket{j,m}\bra{j,m} 
\end{eqnarray}
since $H_1 = H_1^\dagger$ and $H_1Z_1H_1=X_1$.  Expanding $e^{X_1\alpha}$, this equation can be rewritten as
\begin{eqnarray}
\rhoth^\prime &=& \frac{1}{\pfun} \Big[\Big(\cosh\alpha\Big) I_1 + \Big(\sinh\alpha\Big) X_1\Big] \nonumber \\ 
&& \otimes \sum_{j,m} e^{-2m\alpha} \ket{j,m}\bra{j,m} \nonumber \\
&=& \frac{1}{\pfun} \Big[\cosh\alpha\Big(\ket{0}\bra{0}+\ket{1}\bra{1}\Big) \\ 
&+& \sinh\alpha \Big(\ket{0}\bra{1}+\ket{1}\bra{0}\Big)\Big] \otimes\sum_{j,m} e^{-2m\alpha}\ket{j,m}\bra{j,m} \nonumber
\,.
\end{eqnarray}
In the last line, we have expanded $I_1$ and $X_1$ in the computational basis of the first qubit.

\begin{figure*}[ht]
\centerline{
\subfigure[] {\label{fig:uchOneNeg500}
\includegraphics[scale=0.45]{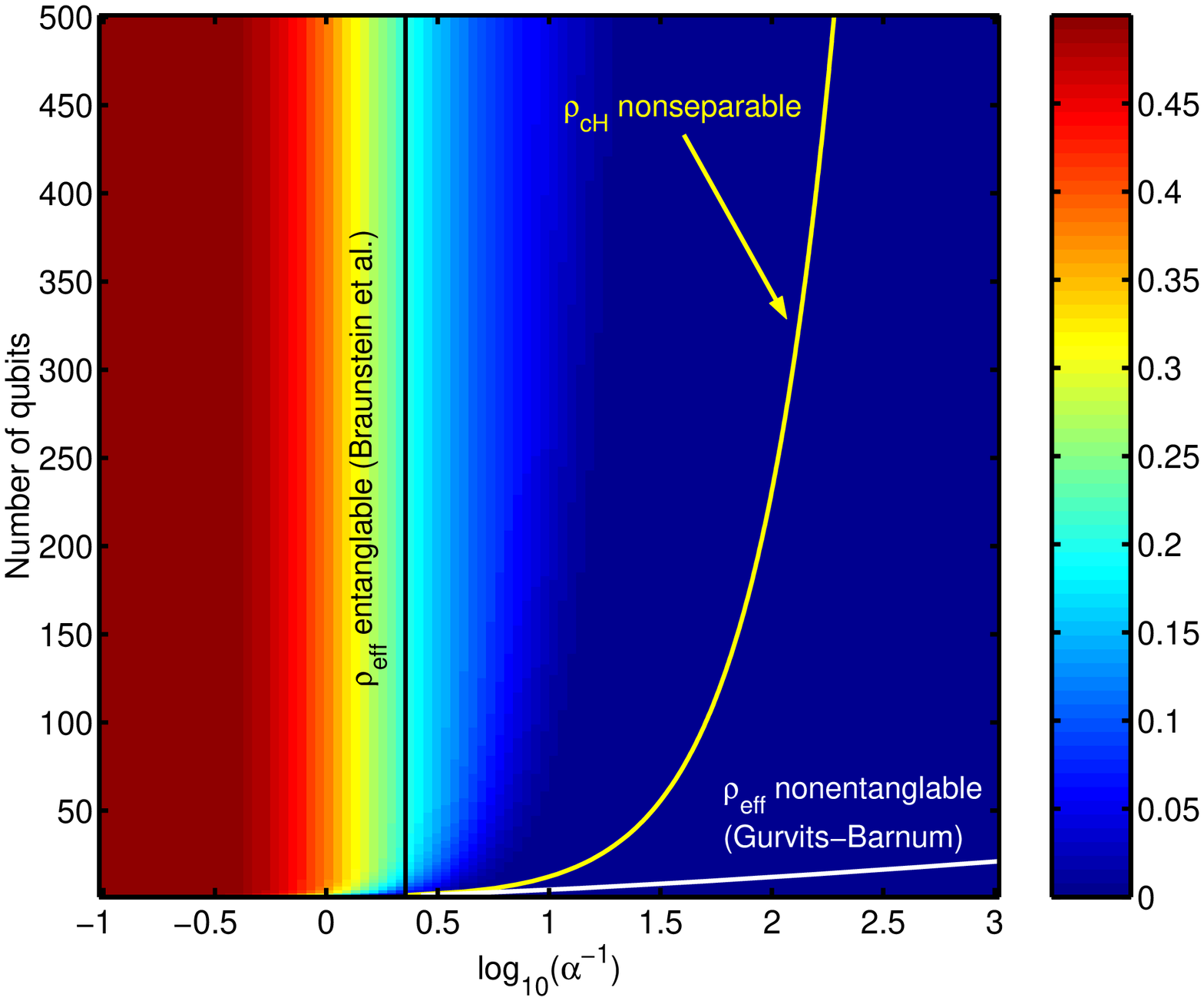}}
\subfigure[] {\label{fig:uchOnelogNeg500}
\includegraphics[scale=0.45]{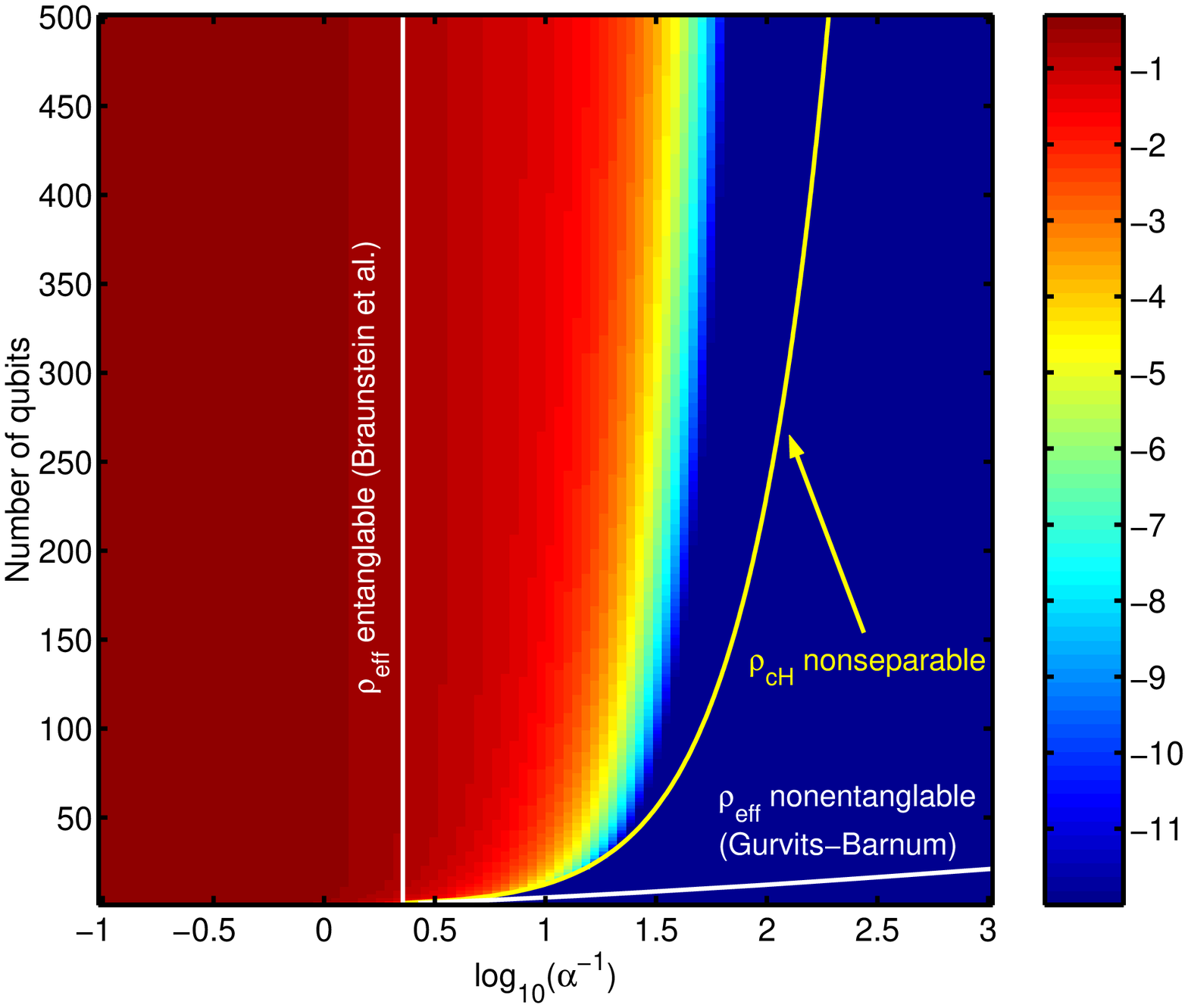}}}
\caption[short list entry]{(color) Negativity (a) and logarithm of negativity (b) of CNOT-Hadamard-transformed thermal states under $\{1,N-1\}$ bipartite split in $N-\alpha$ parameter space.  The logarithm of negativity is calculated as $\log_{10}\left[\negm(\rhoch)+10^{-12}\right]$.  Transformed thermal states $\rhoch$ are nonseparable in the parameter space left of the bound marked ``$\rhoch$ nonseparable'' (Eq.~(\ref{eq:uchboundone})).  Effective pure states are entanglable in parameter space left of the Braunstein \textit{et al.}~bound (Eq.~(\ref{eq:braunent2})) and are nonentanglable in the parameter space beneath the Gurvits-Barnum bound (Eq.~(\ref{eq:gursep2})).}
\label{fig:uchOnefig}
\end{figure*}

To obtain the full transformed state $\rhoch$, we apply the collective CNOT.  Since $\ucnot\ket{0}\ket{j,m} = \ket{0}\ket{j,m}$ and $\ucnot\ket{1}\ket{j,m} = \ket{1}\ket{j,-m}$, we see that $\rhoch$ is composed of two-dimensional subspaces spanned by $\ket{0}\ket{j,m}$ and $\ket{1}\ket{j,-m}$:
\begin{equation}
\rhoch = \bigoplus_{j,m}
\frac{e^{-2m\alpha}}{\pfun}
\lb
\begin{matrix}
\cosh\alpha & \sinh\alpha \\
\sinh\alpha & \cosh\alpha 
\end{matrix}
\rb
\,.
\end{equation}

Next, we calculate the partial transposed state $\rhoch^{T_A}$ with respect to the $\{1,N-1\}$ split.  The partial transpose solely mixes the states in the subspace of $\ket{0}\ket{j,\pm m}$ and $\ket{1}\ket{j,\pm m}$, giving  
\begin{equation}
\rhoch^{T_A} = \bigoplus_{j,m}
\frac{1}{\pfun}
\lb
\begin{matrix}
e^{-2m\alpha}\cosh\alpha & e^{2m\alpha}\sinh\alpha \\
e^{2m\alpha}\sinh\alpha & e^{-2m\alpha}\cosh\alpha 
\end{matrix}
\rb
\,.
\end{equation}
Each term of the above sum has eigenvalues
\begin{equation}
\lambda_\pm(m) = \frac{e^{-2m\alpha}\cosh\alpha\pm e^{2m\alpha}\sinh\alpha}{\pfun}
\label{eq:eig1n-1}
\,.
\end{equation}

The largest possible $m$-value is $(N-1)/2$, so the minimum eigenvalue of $\rhoch^{T_A}$ is
\begin{equation}
\lmin = \frac{e^{-(N-1)\alpha}\cosh\alpha - e^{(N-1)\alpha}\sinh\alpha}{\pfun}
\,.
\label{eq:mineig1n-1}
\end{equation}
When $\lmin$ is negative, $\rhoch$ is entangled according to the negative partial transpose test.  Since $\lmin$ is monotonic in $N$ and $\alpha$, the equality $\lmin=0$ provides a bound on $\rhoch$ being nonseparable:
\begin{equation}
e^{2(N-1)\alpha}\tanh\alpha > 1
\,.
\label{eq:uchboundone}
\end{equation}

As Eq.~(\ref{eq:eig1n-1}) specifies all the eigenvalues of $\rhoch^{T_A}$, we also calculate the negativity by summing $\lambda_\pm$ over all $j$ and $m$-values:
\begin{eqnarray}
\negm(\rhoch) &=& \frac{-1+\sum_{j,m}\Big(|\lambda_+(m)|+|\lambda_-(m)|\Big)}{2} \nonumber \\
&=& -\frac{1}{2}+\frac{1}{2}\sum_j \lp \begin{array}{c} 2\jmax \\ \jmax-j \end{array} \rp\frac{2j+1}{\jmax+j+1} \nonumber \\
&& \times \sum_{m=-j}^j \Big(|\lambda_+(m)| + |\lambda_-(m)|\Big)
\,.
\end{eqnarray}
Here the maximum $j$-value is $\jmax = (N-1)/2$.  The factor in front of the sum over $m$ is the multiplicity of $j$, which is equal to the number of states with $m = j$ minus the number of states with $m = j+1$.  For $N-1$ even, the range of $j$ is $0,1,\ldots\jmax$, whereas for $N-1$ odd, the range of $j$ is $1/2,3/2,\ldots\jmax$.

Fig.~\ref{fig:uchOneNeg500} shows the negativity plotted up to $N$=500.  Note that the horizontal axis is plotted in units of $\log_{10}(\alpha^{-1})$ such that smaller polarization (weaker magnetic fields and higher temperatures) lies towards the right.  As expected, the minimum $\alpha$ required for an entangled thermal state decreases with $N$.  However, the slope of negativity around this $\alpha$ becomes increasingly shallow.  This behavior is seen most clearly in Fig.~\ref{fig:uchOnelogNeg500} where we have plotted the logarithm of negativity~\footnote{This quantity should not be confused with the logarithmic negativity defined in Ref.~\cite{Vidal02a}.} to accentuate small variations from $\negm = 0$.

\subsection{NPT bound for $\{N/2, N/2\}$ bipartite split}

The values of $N$ and $\alpha$ for which the minimum eigenvalue of $\rhoch^{T_A}$ vanishes give a bound on entangled $\rhoch$.  In what follows, we show that it is possible to find an analytic expression for the minimum eigenvalue in terms of $N$ and $\alpha$.  

Numerical calculations reveal that at fixed $N$, the minimum eigenvalue $\lmin$ appears to correspond to one of two eigenvectors depending on whether $\alpha$ is above or below a transition value.  Based on this evidence, we conjecture that the eigenvector corresponding to the minimum eigenvalue is 
\begin{equation}
\ket\vmin = \left\{ \begin{array}{c} \ket{v_-},~\alpha < \alphatr \\
\ket{v_+},~\alpha \geq \alphatr \end{array} \right.
\label{eq:vmin} 
\end{equation}
with $\alphatr$ being the polarization where the transition in eigenvectors occurs.  The eigenvectors are given by
\begin{eqnarray}
\ket{v_-} &=& \frac{1}{\sqrt{2}}\Big(\ket{2^{N-1}-1} - \ket{2^{N-1}}\Big) \label{eq:v-} \\
\ket{v_+} &=& \frac{1}{\sqrt{2}}\Big(\ket{2^{N/2}-1} - \ket{2^N - 2^{N/2}}\Big) \label{eq:v+} 
\end{eqnarray}
in the computational basis.  The state labels are understood in binary notation.  We numerically verified Eqs.~(\ref{eq:vmin})-(\ref{eq:v+}) up to $N=10$. 

We now obtain an analytical formula for the hypothesized $\lmin$.  Using the relation $\rhoch^{T_A}\ket{v_\pm} = \lambda_\pm \ket{v_\pm}$, we find
\begin{eqnarray}
\lambda_- &=& \braket{2^{N-1}-1}{\rhoch^{T_A}}{2^{N-1}-1} \nonumber \\
&& - \braket{2^{N-1}-1}{\rhoch^{T_A}}{2^{N-1}} \label{eq:mineighalf-} \\
\lambda_+ &=& \braket{2^{N/2}-1}{\rhoch^{T_A}}{2^{N/2}-1} \nonumber \\
&& - \braket{2^{N/2}-1}{\rhoch^{T_A}}{2^N-2^{N/2}} \label{eq:mineighalf+} 
\,.
\end{eqnarray}
The minimum eigenvalue is $\lmin = \mathrm{min}(\lambda_-,\lambda_+)$.

\begin{table*}[t]
\begin{center}
\begin{tabular}{l@{\extracolsep{1cm}}l}\hline\hline 
\textbf{Matrix element} & \textbf{Equivalent expression} \\ \hline 
{\rule[-2.4ex]{0pt}{6.0ex}
$\braket{2^{N-1}-1}{\rhoch^{T_A}}{2^{N-1}-1}$} & $\frac{1}{2} \Big(\braket{2^{N-1}-1}{\rhoth}{2^{N-1}-1} + \braket{2^N-1}{\rhoth}{2^N-1} \Big)$ \\ 
{\rule[-2.4ex]{0pt}{6.0ex}
$\braket{2^{N-1}-1}{\rhoch^{T_A}}{2^{N-1}}$} & $\frac{1}{2} \Big(\braket{2^{N-1}-2^{N/2}}{\rhoth}{2^{N-1}-2^{N/2}} - \braket{2^N-2^{N/2}}{\rhoth}{2^N-2^{N/2}} \Big)$ \\ 
{\rule[-2.4ex]{0pt}{6.0ex}
$\braket{2^{N/2}-1}{\rhoch^{T_A}}{2^{N/2}-1}$} & $\frac{1}{2} \Big(\braket{2^{N/2}-1}{\rhoth}{2^{N/2}-1} + \braket{2^{N/2}+2^{N-1}-1}{\rhoth}{2^{N/2}+2^{N-1}-1} \Big)$ \\ 
{\rule[-2.4ex]{0pt}{6.0ex}
$\braket{2^{N/2}-1}{\rhoch^{T_A}}{2^N-2^{N/2}}$} & $\frac{1}{2} \Big(\braket{0}{\rhoth}{0} + \braket{2^{N-1}}{\rhoth}{2^{N-1}} \Big)$  \\ \hline\hline
\end{tabular}
\end{center}
\caption{Formulas for several matrix elements of $\rhoch^{T_A}$ given in terms of the diagonal elements of $\rhoth$.  These expressions are needed to derive an analytical formula for the minimum eigenvalue of $\rhoch^{T_A}$ under the $\{N/2,N/2\}$ bipartite split.}
\label{table:mineigterms}
\end{table*}

The matrix elements in Eqs.~(\ref{eq:mineighalf-}) and~(\ref{eq:mineighalf+}) are calculated in the following manner.  Observe that a given matrix element $\bra{i}\rhoch^{T_A}\ket{j} = \bra{k}\rhoch\ket{\ell}$ where $\ket{k}\bra{\ell} = (\ket{i}\bra{j})^{T_A}$.  Since $\uch$ is sparse and $\rhoth$ is diagonal in the computational basis, $\bra{k}\rhoch\ket{\ell}$ may be expressed as a simple sum of diagonal entries in $\rhoth$.  Formulas for the relevant matrix elements of $\rhoch^{T_A}$ are given in Table~\ref{table:mineigterms}.

Substituting the results from Table~\ref{table:mineigterms} into Eqs.~(\ref{eq:mineighalf-})-(\ref{eq:mineighalf+}) and applying Eq.~(\ref{eq:thstate-diag}), we obtain%
\begin{eqnarray}
\lambda_- &=& \frac{e^\alpha (e^{-N\alpha}\cosh\alpha-\sinh\alpha)}{\pfun} \label{eq:lambda-} \\
\lambda_+ &=& \frac{e^{-\alpha}(\cosh\alpha-e^{N\alpha}\sinh\alpha)}{\pfun} \label{eq:lambda+} 
\,.
\end{eqnarray}

The transition value $\alphatr$ is determined by the condition $\lambda_+-\lambda_- = 0$ and can be calculated by solving the relation
\begin{equation}
e^{N\alphatr}\tanh\alphatr = 1 
\label{eq:alphahalf}
\,.
\end{equation}

The expressions for $\lambda_\pm$ show that Eq.~(\ref{eq:alphahalf}) is also a constraint for $\lmin = 0$.  Therefore, the bound on nonseparable $\rhoch$ can be specified by
\begin{equation}
e^{N\alpha}\tanh\alpha > 1 \label{eq:uchboundhalf}
\,.
\end{equation}

\subsection{Discussion}

\begin{figure}[t]
\includegraphics[width=0.45\textwidth]{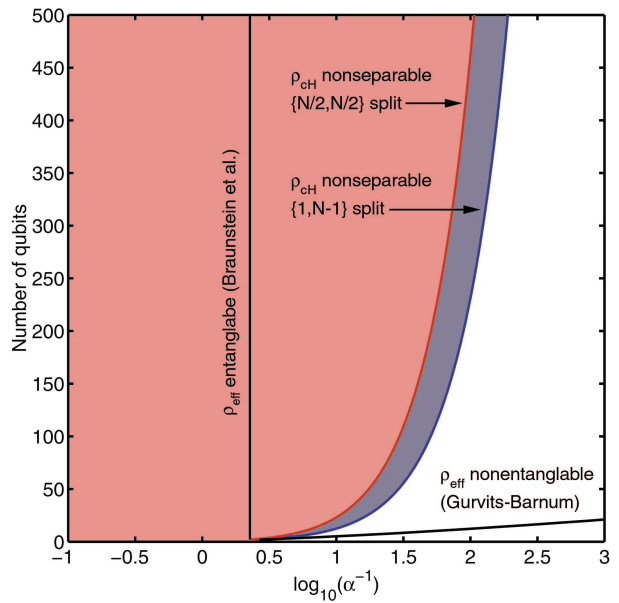}
\caption[short list entry]{(color) Bounds on nonseparable CNOT-Hadamard-transformed thermal states under $\{1, N-1\}$ bipartite split (Eq.~(\ref{eq:uchboundone})) and $\{N/2, N/2\}$ bipartite split (Eq.~(\ref{eq:uchboundhalf})).  The light red region marks parameter space where $\rhoch$ is nonseparable under the $\{N/2,N/2\}$ split.  The union of the light red and light blue regions shows where $\rhoch$ is nonseparable under the $\{1,N-1\}$ split.  Effective pure states are entanglable in parameter space left of the Braunstein \textit{et al.}~lower bound (Eq.~(\ref{eq:braunent2})) and are nonentanglable in the parameter space beneath the Gurvits-Barnum bound (Eq.~(\ref{eq:gursep2})).}
\label{fig:splitcomp}
\end{figure}

We compare the NPT bounds on the entanglability of thermal states with comparable bounds on effective pure states in $N$-$\alpha$ parameter space.  Fig.~\ref{fig:splitcomp} shows the bounds on nonseparable $\rhoch$ for the $\{1,N-1\}$ and $\{N/2,N/2\}$ split in $N$-$\alpha$ space.  In both cases, the thermal state bounds include more parameter space than the Braunstein \textit{et al.}~lower bound, although the $\{1,N-1\}$ split gives the tighter constraint.  Neither thermal state bound succeeds in capturing parameter space where $\rhoeff$ is nonentanglable.

Numerically solving Eq.~(\ref{eq:uchboundone}), we find that for polarization $\alpha = 4 \times 10^{-5}$, we require at least 126,584 qubits to achieve entanglement for $\uch$ transformed thermal states.  This number is clearly impractical, but it dramatically improves upon the Schulman-Vazirani value of $6.3 \times 10^8$ qubits from Sec.~\ref{sec:exp}.


\section{Bounds on the separability and distillability of Bell-transformed thermal states} \label{sec:bounds}

We now apply a more general procedure to derive bounds on the separability and distillability of Bell-transformed thermal states.  Unlike the methods of Sec.~\ref{sec:npt}, this procedure allows us to find bounds on the entanglement of thermal states transformed by any Bell unitary under any bipartite split.  The main idea is to take a state $\rho^\prime$ with known entanglement properties and find a random local unitary operation that transforms $\ubell\rhoth\ubell^\dagger \mapsto \rho^\prime$.  Because such an operation cannot increase the entanglement of the initial state, the entanglement of $\rho^\prime$ gives a lower bound on the entanglement of Bell-transformed thermal states.

A candidate state for $\rho^\prime$ is provided by a family of mixed Bell states $\rho_N$ whose entanglement has been classified by D\"ur and Cirac~\cite{Dur00a}.  We first review their formalism and then apply it to Bell-transformed thermal states to find bounds on their separability and distillability.  These results give tighter bounds than the direct calculations of Sec.~\ref{sec:npt}.

\subsection{D\"ur-Cirac classification of entanglement in special mixed Bell states} 

\subsubsection{Formalism}

D\"ur and Cirac consider a special family of mixed Bell states parameterized by the number of qubits $N$.  They are defined by
\begin{eqnarray}
\rho_N &=& \lambda_0^+ \ket{\Psi_0^+}\bra{\Psi_0^+} + \lambda_0^- \ket{\Psi_0^-}\bra{\Psi_0^-} \nonumber \\
&& + \sum_{j=1}^{2^{(N-1)}-1} \lambda_j \Big(\ket{\Psi_j^+}\bra{\Psi_j^+} + \ket{\Psi_j^-}\bra{\Psi_j^-}\Big) \label{eq:rhon}
\end{eqnarray}
where the generalized Bell states are given by
\begin{equation}
\ket{\Psi_j^\pm} = \frac{1}{\sqrt{2}}\Big(\ket{0}\ket{j}\pm\ket{1}\ket{\bar{j}}\Big) \label{eq:genbell}
\end{equation}
in the computational basis.  The state of the first qubit and the state of the latter $N-1$ qubits are explicitly separated~\footnote{D\"ur and Cirac separate out the state of the first $N-1$ qubits and the last qubit.  We choose a slightly different formulation for convenience in matching our previous formulas.}.  Here $\ket{\bar{j}} \equiv \ket{2^{N-1}-j-1}$ is the bit-flipped version of $\ket{j}$, and $1 \leq j < 2^{N-1}-1$.  The parameters $\lambda_0^\pm$ are chosen such that $\lambda_0^+ - \lambda_0^- \geq 0$.  

We specify a bipartite split with a nonnegative integer $k$, which when expressed in binary, labels the qubits that are in party $A$ with 1s and the qubits in party $B$ with 0s.  For example, if $k = 010110$, then the second, fourth, and fifth qubits are in party $A$, and the first, third, and sixth qubits are in party $B$.  With no loss in generality, we require the first qubit to always be in party $B$, i.e. the most significant bit is always 0.  Therefore, the possible bipartite splits lie in the range $1 \leq k \leq 2^{N-1}-1$.

The family of states $\rho_N$ has particularly elegant conditions on its partial transposition.  A straightforward calculation shows that $\rho_N$ under bipartite split $k$ has the properties:
\begin{eqnarray}
\Delta &\leq& 2\lambda_k \Rightarrow \rho_N~\mathrm{has~PPT} \label{eq:rhonppt} \\
\Delta &>& 2\lambda_k \Rightarrow \rho_N~\mathrm{has ~NPT} \label{eq:rhonnpt}
\end{eqnarray}  
where we have defined $\Delta \equiv \lambda_0^+ - \lambda_0^-$.

More importantly, we can characterize the full separability and distillability of $\rho_N$.  D\"ur and Cirac proved the following:
\begin{enumerate}
\item Consider all possible bipartite splits of an $N$ qubit system.  If and only if each of these splits has PPT, then $\rho_N$ is fully separable, i.e. separable under any partition of the system.
\item Consider all possible bipartite splits of an $N$ qubit system.  If and only if each of these splits has NPT, then $\rho_N$ is fully distillable, i.e. a maximally entangled pair can be distilled from any two particles in the system.
\end{enumerate}
Thus the PPT and NPT conditions give bounds on the separability and distillability of $\rho_N$:
\begin{eqnarray}
\Delta &\leq& 2~\mathrm{min}_{\{k\}} (\lambda_k) \Rightarrow \mbox{$\rho_N$ fully separable} \label{eq:rhonsep} \\
\Delta &>& 2~\mathrm{max}_{\{k\}} (\lambda_k) \Rightarrow \mbox{$\rho_N$ fully distillable} \label{eq:rhondist}
\end{eqnarray}
where the minimum and maximum are taken over all possible bipartite splits $k$.

The bound on full distillability is significant because it specifies the regions in parameter space where \textit{useful} entanglement can be obtained in the form of maximally entangled pairs.  

\subsubsection{Bound on the entanglability of effective pure states}

As D\"ur and Cirac discuss~\cite{Dur00a}, the fully distillable criterion gives a simple bound on the entanglability of effective pure states that is tighter than the Braunstein \textit{et al.}~lower bound in Eq.~(\ref{eq:braunent}).  

Suppose we apply a unitary to the effective pure state that transforms $\ket{0}$ to the maximally entangled Bell state $\ket{\Psi_0^+}$, yielding a new state
\begin{equation}
\rhoeff^\prime = \frac{1-\epsilon}{d} I_d + \epsilon\ket{\Psi_0^+}\bra{\Psi_0^+} \label{eq:tf-rhoeff}
\end{equation}
that fits the form of $\rho_N$.  It is easy to see that $\lambda_0^+ = (1-\epsilon)/d + \epsilon$ and $\lambda_0^- = \lambda_j = (1-\epsilon)/d$.  Therefore, $\Delta = \epsilon$.  

Eq.~(\ref{eq:rhondist}) shows that full distillability~\footnote{According to Eq.~(\ref{eq:rhonsep}), we also obtain a condition for full separability of the transformed effective pure state.  However, this criterion is relevant to a specific group of unitaries and therefore does not give as much information as the Braunstein~\textit{et al.}~and Gurvits-Barnum separable bounds.} of the transformed $\rhoeff$ requires
\begin{equation}
\epsilon > \frac{1}{1+2^{N-1}} \label{eq:dcdist-rhoeff}
\,.
\end{equation}

Substituting Eq.~(\ref{eq:eps-rhoeff}), we obtain a bound on the entanglability of effective pure states in NMR parameter space:
\begin{equation}
\alpha > -\frac{1}{2}\ln\lb\lp\frac{2+2^N}{3}\rp^{1/N}-1\rb
\,.
\label{eq:durent}
\end{equation}

\subsection{Application to Bell-transformed thermal states} 

\subsubsection{Random local unitary operation mapping Bell-transformed thermal states to $\rho_N$} 

The thermal state is diagonal in the computational basis, so any Bell-transformed thermal state has the form
\begin{eqnarray}
\rhob &=& \lambda_0^+ \ket{\Psi_j^+}\bra{\Psi_j^+} + \lambda_0^- \ket{\Psi_j^-}\bra{\Psi_j^-} \\
&& + \sum_{j=1}^{2^{(N-1)}-1} \Big(\lambda_j^+ \ket{\Psi_j^+}\bra{\Psi_j^+} + \lambda_j^- \ket{\Psi_j^-}\bra{\Psi_j^-}\Big) \nonumber
\,.
\end{eqnarray}
The state is diagonal in the generalized Bell basis $\left\{\ket{\Psi_j^\pm}\right\}$, but is not of the form $\rho_N$ because generally $\lambda_j^+ \neq \lambda_j^-$.

Suppose we can find a random local unitary operation $\eunit$ such that $\eunit(\rhob)$ has form $\rho_N$.  The action of any random local unitary operation on an arbitrary $\rho$ can always be expressed as a probabilistic mixture of unitaries $U_i$:
\begin{equation}
\eunit(\rho) = \sum_i p_i U_i \rho U_i^\dagger \label{eq:randomunit} 
\end{equation}
where $p_i$ is the probability of applying the unitary $U_i$.  All the unitaries $U_i$ are local, meaning that they act on each qubit separately.  Therefore the entanglement of each state $U_i \rho U_i^\dagger$ cannot be larger than the entanglement of the original state $\rho$.  Moreover, the quantum operation $\eunit(\rho)$ can be interpreted as a convex combination of transformed thermal states.  Thus the entanglement of $\eunit(\rho)$ must be less than or equal to the entanglement of $\rho$ itself; a random mixture of local unitaries \textit{cannot increase} entanglement.  It follows that the entanglement of $\eunit(\rhob)$ must bound the entanglement of $\rhob$. 

Now we describe a particular random local unitary operation $\mathcal{E}_\phi$ that performs the desired transformation.  Consider the following procedure:
\begin{enumerate}
\item Start with the Bell-transformed thermal state $\rhob$.
\item Apply mixing operation $R=\bigotimes_{i=1}^N R_i$ where $R_i$ multiplies the $i$th qubit by a random phase $\phi_i$ if the qubit has state $\ket{0}$ and $R_i$ does nothing otherwise.  We assume that $\phi_i$ is uniformly distributed over $[-\pi, \pi]$ subject to the constraint $\sum_i \phi_i = 2\pi$.  This requirement is chosen so that $R\ket{\Psi_0^\pm} = \ket{\Psi_0^\pm}$.
\item Average $R\rhob$ over all $\phi_i$.
\end{enumerate}

Let us verify that $\mathcal{E}_\phi$ produces a state of form $\rho_N$.  First, we calculate the effect of $R$ on each generalized Bell state:
\begin{eqnarray}
\ket{\chi_j^\pm} &=& R\ket{\Psi_j^\pm} \nonumber \\
                 &=& \frac{1}{\sqrt{2}}\Big( e^{i\theta_j}\ket{0}\ket{j} \pm e^{-i\theta_j}\ket{1}\ket{\bar{j}}\Big)
\end{eqnarray}
where $\theta_j$, the phase of $\ket{0}\ket{j}$ due to the action of $R$, is given by
\begin{equation}
\theta_j = \sum_{\{i|f(i,j)=1\}} \phi_i
\,.
\end{equation}
The function $f$ is unity only if the $i$th binary digit of $j$ is one.  Note that $\ket{1}\ket{\bar{j}}$ acquires a phase of $-\theta_j$ because it is the bit-flipped version of $\ket{0}\ket{j}$.

Next, we find the state that results when $R$ acts on the entire Bell-transformed thermal state:
\begin{equation}
R \rhob R^\dagger = \sum_{j=0}^{2^{(N-1)}-1} \Big(\lambda_j^+ \ket{\chi_j^+}\bra{\chi_j^+} + \lambda_j^- \ket{\chi_j^-}\bra{\chi_j^-}\Big) \label{eq:Rrhotf}
\end{equation} 
where
\begin{eqnarray}
\ket{\chi_j^\pm}\bra{\chi_j^\pm} &=& \frac{1}{2}\Big [\Big(\ket{0}\ket{j}\bra{0}\bra{j} + \ket{1}\ket{\bar{j}}\bra{1}\bra{\bar{j}}\Big) \label{eq:oproduct} \\
&&{} \pm \Big( e^{2i\theta_j}\ket{0}\ket{j}\bra{1}\bra{\bar{j}}+e^{-2i\theta_j}\ket{1}\ket{\bar{j}}\bra{0}\bra{j}\Big)\Big ] \nonumber 
\,.
\end{eqnarray}

When we average Eq.~(\ref{eq:Rrhotf}) over $\phi_i$, the last two terms in Eq.~(\ref{eq:oproduct}) vanish for all $j \neq 0$.  When $j=0$, $\mathcal{E}_\phi$ has no effect as $\theta_0$ is constrained to be $2\pi$.  The final state after application of $\mathcal{E}_\phi$ is consequently
\begin{eqnarray}
\mathcal{E}_\phi(\rhob) &=& \lambda_0^+ \ket{\Psi_0^+}\bra{\Psi_0^+} + \lambda_0^- \ket{\Psi_0^-}\bra{\Psi_0^-} \\
&& + \sum_{j=1}^{2^{(N-1)}-1} \frac{\lambda_j^+ + \lambda_j^-}{2} \Big(\ket{\Psi_j^+}\bra{\Psi_j^+} + \ket{\Psi_j^-}\bra{\Psi_j^-}\Big) \nonumber 
\,,
\end{eqnarray}
which is indeed of form $\rho_N$.  Matching this expression with Eq.~(\ref{eq:rhon}), we obtain 
\begin{equation}
\lambda_j = \frac{\lambda_j^+ + \lambda_j^-}{2} \label{eq:lj}
\,.
\end{equation}
The effect of $\mathcal{E}_\phi$ is to average the eigenvalues $\lambda_j^\pm$.

We emphasize that $\mathcal{E}_\phi$ is not the only possible operation that produces the needed transformation.  Random local unitary operations generally \textit{decrease} the entanglement of the system.  We desire a procedure that preserves as much entanglement as possible.  In the next subsection, we will make a few remarks about the effectiveness of $\mathcal{E}_\phi$ in this regard.

\subsubsection{Separability and distillability of $\uch$ transformed thermal states}

We now calculate separability and distillability bounds on Bell-transformed thermal states.  First, we examine the case where the Bell unitary is chosen to be the CNOT-Hadamard transformation $\uch$.

Fig.~\ref{fig:uchufan} shows that $\uch$ maps computational basis states to Bell states in the following manner:
\begin{eqnarray}
\ket{i} &\mapsto& \ket{\Psi_i^+},~0 \leq i < 2^{N-1}-1 \nonumber \\
\ket{i} &\mapsto& \ket{\Psi_{i-2^{(N-1)}}^-},~2^{N-1} \leq i < 2^N \label{eq:uch-map}
\,.
\end{eqnarray}
Inserting Eq.~(\ref{eq:thstate-diag}), we find
\begin{eqnarray}
\lambda_j^+ &=& \bra{\Psi_j^+}\rhoch\ket{\Psi_j^+} = \frac{e^{[N-2w(j)]\alpha}}{\pfun} \\
\lambda_j^- &=& \bra{\Psi_j^-}\rhoch\ket{\Psi_j^-} = \frac{e^{[N-2w(j)-2]\alpha}}{\pfun} 
\,.
\end{eqnarray}

Then we apply the random local unitary operation $\mathcal{E}_\phi$ and find that $\mathcal{E}_\phi(\rhoch)$, when expressed in the form of $\rho_N$, has parameters
\begin{eqnarray}
\Delta &=& \frac{1}{\pfun}~e^{N\alpha}\Big(1-e^{-2\alpha}\Big) \\
\lambda_k &=& \frac{1}{2\pfun}~e^{[N-2w(k)]\alpha}\Big(1+e^{-2\alpha}\Big)
\,.
\end{eqnarray}

Inserting the above into Eqs.~(\ref{eq:rhonppt}) and~(\ref{eq:rhonnpt}), we establish PPT and NPT conditions on $\rhoch$ under bipartite split $k$:
\begin{eqnarray}
\tanh\alpha &\leq& e^{-2w(k)\alpha} \Rightarrow \rhoch~\mathrm{has~PPT} \label{eq:uchppt} \\ 
\tanh\alpha &>& e^{-2w(k)\alpha} \Rightarrow \rhoch~\mathrm{has~NPT} \label{eq:uchnpt}
\,.
\end{eqnarray}

From the above conditions, the entanglement of $\rhob$ is constrained by the bounds 
\begin{eqnarray}
\tanh\alpha &\leq& e^{-2(N-1)\alpha} \Rightarrow \rhob~\mathrm{fully~separable} \label{eq:uch-sepbound} \\
\tanh\alpha &>& e^{-2\alpha} \Rightarrow \rhob~\mathrm{fully~distillable} \label{eq:uch-distbound}
\,.
\end{eqnarray}
The separable bound corresponds to the $\{1,N-1\}$ split, and the distillable bound corresponds to the $\{N-1,1\}$ split.

\subsubsection{Separability and distillability of $\uch\ufan$ transformed thermal states} 

We turn to the case where the Bell unitary is chosen to be $\uch\ufan$.  This unitary yields tighter bounds compared to $\uch$, as we now show.

The tightness of a bound derived from D\"ur-Cirac formalism increases with the size of $\Delta$.  In the case where we transform $\rhoth$ with $\uch$, we effectively set $\lambda_0^+ = e^{N\alpha}/\pfun$ and $\lambda_0^- = e^{(N-2)\alpha}/\pfun$, giving a gap $\Delta$ that scales as $2^{-N}$.  However, a larger gap may be achieved.

We can shuffle the unitary mapping between computational basis states and Bell states to obtain the values of $\lambda_0^\pm$ and $\lambda_j$ that maximize $\Delta$.  The largest possible gap occurs when we select $\lambda_0^\pm = e^{\pm N\alpha}/\pfun$, which yields  
\begin{equation}
\Delta = \frac{2\sinh(N\alpha)}{\pfun} 
\,.
\label{eq:delta-uchufan}
\end{equation}
This choice scales as $N2^{-N}$.

The new gap can be realized with mapping
\begin{eqnarray}
\ket{i} &\mapsto& \ket{\Psi_i^+},~0 \leq i < 2^{N-1}-1 \nonumber \\
\ket{i} &\mapsto& \ket{\Psi_{2^N-i-1}^-},~2^{N-1} \leq i < 2^N \label{eq:uchufan-map}
\,.
\end{eqnarray} 
This mapping is exactly produced by the Bell unitary $\uch\ufan$, with the permutation matrix $\ufan$ shuffling the old mapping in Eq.~(\ref{eq:uch-map}) to give Eq.~(\ref{eq:uchufan-map}).

We proceed to find bounds on the separability and distillability of $\rhob$.  Doing the same calculation as before, we have
\begin{equation}
\lambda_k = \frac{1}{\pfun}\cosh\lb\Big(N-2w(k)\Big)\alpha\rb \label{eq:lk-uchufan}
\,.
\end{equation}

Comparing Eqs.~(\ref{eq:delta-uchufan}) and~(\ref{eq:lk-uchufan}), we establish the following conditions on $\rhocf$ under bipartite split $k$:
\begin{eqnarray}
\tanh(k\alpha) &\leq& e^{-2[N-w(k)]\alpha} \Rightarrow \rhocf~\mathrm{has~PPT} \\
\tanh(k\alpha) &>& e^{-2[N-w(k)]\alpha} \Rightarrow \rhocf~\mathrm{has~NPT}
\,.
\end{eqnarray}

\begin{figure*}[t]
\centerline{
\subfigure[] {\label{fig:dcbounds20}
\includegraphics[scale=0.45]{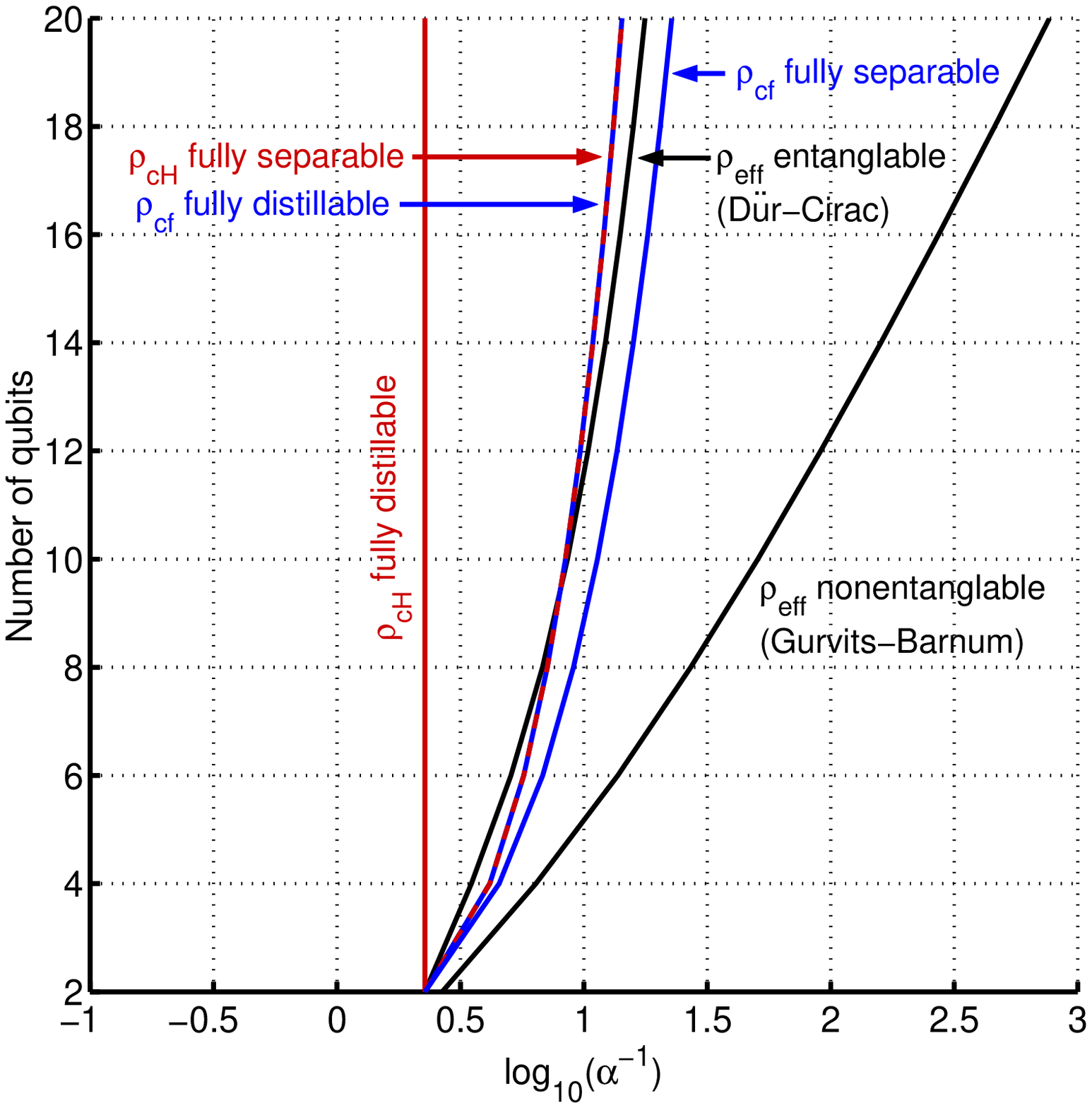}}
~~~~~~\subfigure[] {\label{fig:dcbounds500}
\includegraphics[scale=0.45]{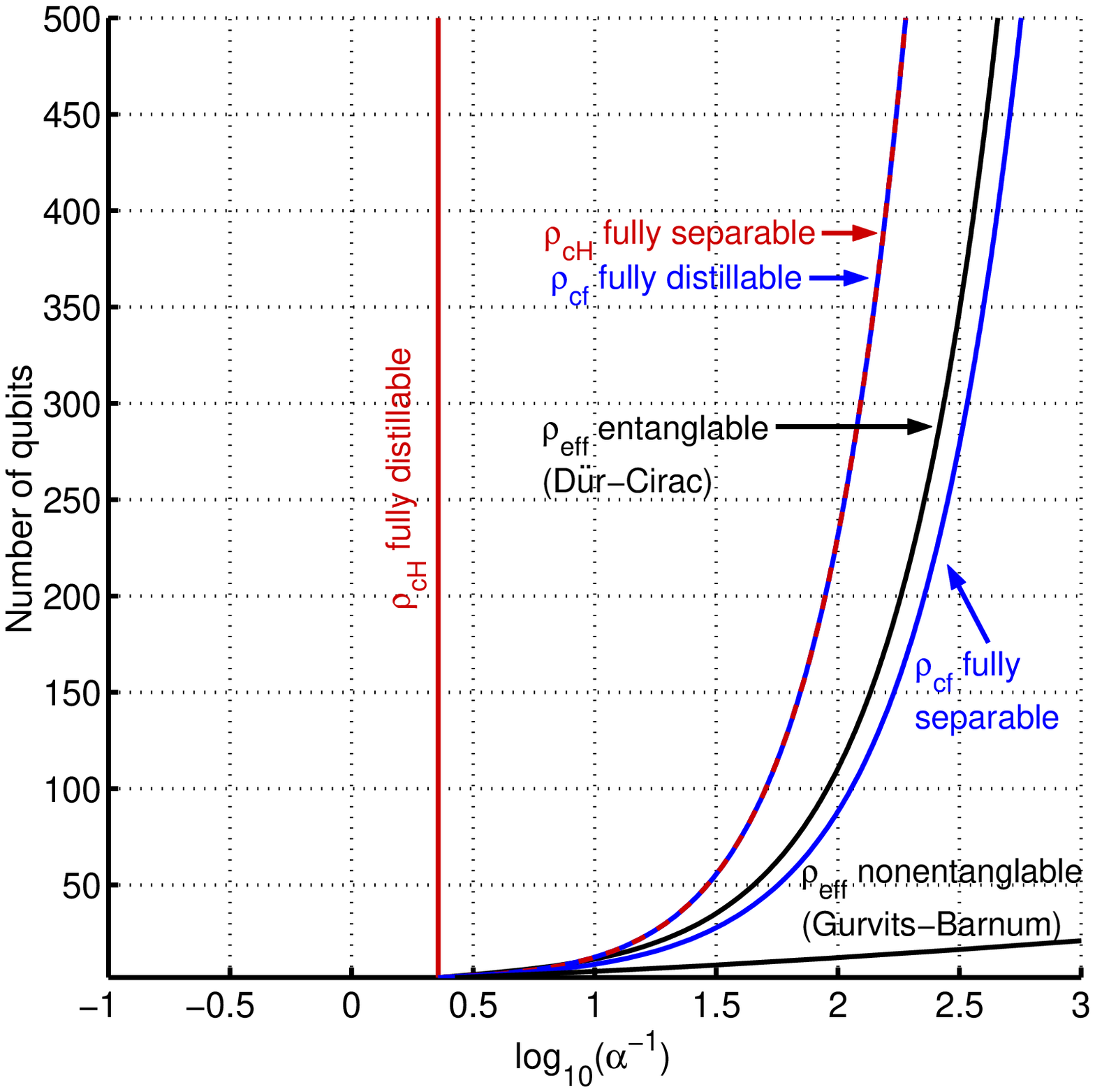}}}
\caption[short list entry]{(color) Bounds on the separability and distillability of Bell-transformed thermal states as derived from D\"ur-Cirac formalism, (a) $2 \leq N \leq 20$ and (b) $2 \leq N \leq 500$: $\rhoch$ fully separable (Eq.~(\ref{eq:uch-sepbound})), $\rhoch$ fully distillable (Eq.~(\ref{eq:uch-distbound})), $\rhocf$ fully separable (Eq.~(\ref{eq:uchufan-sep})), and $\rhocf$ fully distillable (Eq.~(\ref{eq:uchufan-dist})).  Transformed thermal states $\rhoch$ and $\rhocf$ are separable/distillable in the parameter space to the right/left of the appropriate separable/distillable bound.  Effective pure states are nonentanglable in the parameter space beneath the Gurvits-Barnum bound (Eq.~(\ref{eq:gursep2})) and are entanglable in the parameter space left of the D\"ur-Cirac bound (Eq.~(\ref{eq:durent})).}
\label{fig:dcbounds}
\end{figure*}

Minimizing and maximizing the above conditions over all possible bipartite splits, we find that $\rhob$ is fully separable if and only if
\begin{equation}
\sinh(N\alpha) \leq 1 \label{eq:uchufan-sep}
\end{equation}
and that $\rhob$ is fully distillable if and only if
\begin{subequations}
\begin{eqnarray}
\tanh\lb(N-1)\alpha\rb &>& e^{-2\alpha}~\mathrm{or} \label{eq:uchufan-dist1} \\
\tanh\alpha &>& e^{-2(N-1)\alpha} \label{eq:uchufan-dist2}
\,.
\end{eqnarray}
\label{eq:uchufan-dist}
\end{subequations}
Eq.~(\ref{eq:uchufan-sep}) is derived from the $\{N/2,N/2\}$ split whereas Eqs.~(\ref{eq:uchufan-dist1}) and~(\ref{eq:uchufan-dist2}) correspond to the $\{1,N-1\}$ and $\{N-1,1\}$ splits respectively.  These two splits give the same bound because $\lambda_k$ in Eq.~(\ref{eq:lk-uchufan}) is symmetric under the replacement of $w(k)$ by $N-w(k)$.

\subsection{Discussion}

We first evaluate the effectiveness of our random local unitary operation $\mathcal{E}_\phi$.  Observe that the $\{1,N-1\}$ and $\{N/2,N/2\}$ bipartite splits correspond to Hamming weights $w(k)$ of $N-1$ and $N/2$ respectively.  If these weights are inserted into Eq.~(\ref{eq:uchnpt}), we recover the NPT bounds calculated in Eqs.~(\ref{eq:uchboundone}) and~(\ref{eq:uchboundhalf}).  We also numerically computed NPT bounds (up to $N=10$) for the bipartite splits corresponding to the bounds on fully separable and distillable $\rhob$ in Eqs.~(\ref{eq:uch-sepbound}), (\ref{eq:uch-distbound}), (\ref{eq:uchufan-sep}) and (\ref{eq:uchufan-dist}).  The excellent agreement in all cases implies that our procedure $\mathcal{E}_\phi$ does not erase any of the information we obtain from the direct NPT test.

We now examine the behavior of the D\"ur-Cirac derived bounds on fully separable and distillable Bell-transformed thermal states in NMR parameter space.  These results are shown in Fig.~\ref{fig:dcbounds}.  As a comparison, we have also plotted bounds on the entanglability (D\"ur and Cirac) and nonentanglability (Gurvits and Barnum) of effective pure states.  Note that the bound on entanglable effective pure states also implies full distillability of transformed $\rhoeff$.

The $\uch\ufan$ unitary produces tighter bounds than $\uch$, as we expect.  Curiously, the bound on fully distillable $\rhoch$ exactly coincides with the Braunstein \textit{et al.} bound on the nonentanglability of $\rhoeff$ (see Fig.~\ref{fig:splitcomp}), and the bound for fully separable $\rhoch$ exactly matches the bound on fully distillable $\rhocf$.

We are interested in knowing whether our thermal state bounds improve upon the known constraints on effective pure states.  First, we compare the bound on fully distillable $\rhocf$ against the corresponding bound for effective pure states.  The thermal state bound encompasses more parameter space at small $N$, but the effective pure state bound does better for $N > 9$.  Yet thermal states should be at least as entanglable as effective pure states.  The apparent contradiction indicates that while $\uch\ufan$ is the optimal Bell unitary, it is not the optimal unitary in general for entangling thermal states.  Second, we observe that the D\"ur-Cirac derived bounds on separable $\rhob$ are much looser than the Gurvits-Barnum bound on the nonentanglability of $\rhoeff$.  This result again suggests that the family of unitaries $\ubell$ is not optimal.  

We do succeed in finding entangled thermal states outside the D\"ur-Cirac bound on entanglable effective pure states.  Thermal states outside the parameter space of fully separable $\rhob$ have at least one NPT bipartite split, implying that at least one maximally entangled pair can be distilled.  The bound on fully separable $\rhocf$ (Eq.~(\ref{eq:uchufan-sep})) thus gives the tightest bound on entanglable thermal states, scaling as $N \sim 1/\alpha$.  It also captures new parameter space outside of the D\"ur-Cirac bound on entanglable $\rhoeff$.  Assuming the largest realistic initial polarization $\alpha = 4 \times 10^{-5}$, Eq.~(\ref{eq:uchufan-sep}) requires $N \leq 22,035$ for entanglable thermal states.  In comparison, the D\"ur-Cirac bound in Eq.~\ref{eq:durent} gives a constraint $N \leq 50,000$ for entanglable effective pure states in the same parameter space.


\section{Majorization approach to determining entanglability} \label{sec:majorization}

We have seen that the D\"ur-Cirac formalism is a powerful tool for obtaining bounds on the entanglement of Bell-transformed thermal states.  The key is to find a random local unitary operation that maps the Bell-transformed thermal state to a state of form $\rho_N$.  In principle, we can always try to construct the needed quantum operation, but this method is time-consuming and wasteful.  Knowing such a procedure exists is enough.  

Motivated by the connection between entanglement and the mathematical theory of majorization~\cite{Nielsen99a}, we discuss a different approach that uses majorization to address this problem.  We first define majorization and describe two theorems from majorization theory: Uhlmann's theorem and the majorization condition on von Neumann entropy.  Then we explain how these theorems may be used to potentially derive tighter bounds on entanglable thermal states.

\subsection{Formalism}

Majorization is a mathematical relation that states whether one probability distribution is more disordered than another.

Suppose we have two probability distributions that can be described by real vectors of length $d$: $\vec{r} = (r_1, r_2, \ldots r_d)$ and $\vec{s} = (s_1, s_2, \ldots, s_d)$.  Let $\vec{r}^\uparrow$ and $\vec{s}^\uparrow$ be the vectors $\vec{r}$ and $\vec{s}$ with their components arranged in non-decreasing order, that is $r_1^\uparrow \leq r_2^\uparrow \leq \ldots \leq r_d^\uparrow$ and vice versa.  We say ``$\vec{s}$ majorizes $\vec{r}$'' or ``$\vec{r}$ is majorized by $\vec{s}$'' if and only if
\begin{equation}
\sum_{j=1}^k r_j^\uparrow \geq \sum_{j=1}^k s_j^\uparrow 
\end{equation}
for $k = 1,2,\ldots,d$ and with the equality holding for $k=d$.  This relation is denoted by $\vec{r} \prec \vec{s}$.  When it holds, we may think of $\vec{r}$ as being more ``disordered'' than $\vec{s}$~\cite{Nielsen02a}.

In the context of quantum mechanics, we take $\vec{r}$ and $\vec{s}$ to be vectors containing the eigenvalues of density matrices.  Given two density matrices $\rho$ and $\sigma$, we have $\rho \prec \sigma$ if and only if $\vec\lambda(\rho) \prec \vec\lambda(\sigma)$ where $\vec\lambda(\rho)$ is a vector containing the eigenvalues of $\rho$.  This interpretation makes sense because the eigenvalues define a probability distribution on an orthogonal set of states.  Thus, majorization for two density matrices is equivalent to comparing the disorder of their corresponding probability distributions.  

There are two results from majorization theory that are useful for our purposes: Uhlmann's theorem~\cite{Uhlmann70a, Uhlmann71a, Uhlmann72a, Uhlmann73a} and the majorization condition on von Neumann entropy~\cite{Nielsen02a}.

Uhlmann's theorem states that given two Hermitian matrices $A$ and $B$, there exists a random unitary operation $\mathcal{E}$ such that $A = \mathcal{E}(B)$ if and only if $B$ majorizes $A$.  Mathematically, we write 
\begin{equation}
A \prec B \Leftrightarrow \exists~\mathcal{E}~\textrm{s.t.}~A = \mathcal{E}(B) = \sum_i p_i U_i B U_i^\dagger
\,. 
\end{equation}
Note that unlike the operation $\eunit$ of Sec.~\ref{sec:bounds}, the unitaries $U_i$ here need not be local. 

The majorization condition on von Neumann entropy $S$ is the following.  Given two density matrices $\rho$ and $\sigma$, if $\sigma$ majorizes $\rho$, the von Neumann entropy of $\rho$ must be at least as large as the von Neumann entropy of $\sigma$, i.e.  
\begin{equation}
\rho \prec \sigma \Rightarrow S(\rho) \geq S(\sigma) \label{eq:maj-ent}
\,.
\end{equation}
Notice that the majorization relation is merely necessary and not sufficient.  

\subsection{Application to thermal states} 

Now we explain how the formalism we just described may be used to derive bounds on the entanglability of thermal states.

Suppose $\rhoth$ majorizes $\rho^\prime$, a state of form $\rho_N$.  Then Uhlmann's theorem says that $\rho^\prime = \sum_i U_i \rhoth U_i^\dagger$.  The same convexity argument from Sec.~\ref{sec:exp} establishes that the entanglement of $\rho^\prime$ gives a lower bound on the maximum entanglement attainable from thermal states.

We have converted the problem of calculating the entanglement of $U\rhoth U^\dagger$ to finding a density matrix $\rho^\prime$ that is majorized by $\rhoth$.  Supposing that $\rho^\prime$ is well-defined in terms of $N$ and $\alpha$, we can derive analytical bounds on entanglable thermal states.

Unfortunately, this majorization approach is not necessarily simpler.  We need to guess the right form for $\rho^\prime$, a state with $2^{N-1}+1$ parameters.  The only constraint is that $\rho^\prime$ be entangled with respect to at least one bipartite split.  Then we must compare as many as $2^N$ eigenvalues to see if the majorization actually holds.  

However, we can use Eq.~(\ref{eq:maj-ent}) as a first cut criterion before checking that the desired majorization relation holds.  When $S(\rhoth) > S(\rho^\prime)$, it is impossible for $\rhoth$ to majorize $\rho^\prime$.  The von Neumann entropy of the thermal state has the simple formula
\begin{equation}
S(\rhoth) = -N \lb p \log_2(p) + (1-p) \log_2(1-p)\rb
\end{equation}
where $p = (1+\tanh\alpha)/2$.  So, if there is a closed form expression for $S(\rho^\prime$), it is easy to see if Eq.~(\ref{eq:maj-ent}) holds.  This method may save time in searching for the best construction of $\rho^\prime$.  

We propose two different strategies.  To simplify matters, we hold $\alpha$ constant.  The first idea is to guess an analytical form for $\rho^\prime$, such that at least one bipartite split has NPT.  We can derive a formula for $S(\rho^\prime$) and use the majorization condition on von Neumann entropy to find the minimum number of qubits needed for $S(\rhoth) > S(\rho^\prime)$.  We then set $N$ to this minimum value and increment $N$ until the majorization relation $\rho^\prime \prec \rhoth$ holds.  Alternatively, we can attempt to numerically optimize $\rho^\prime$ to be majorized by $\rhoth$ at fixed $N$, increasing $N$ if the optimization fails.  

While we have not yet found a good construction for $\rho^\prime$, we believe this majorization-based approach may be fruitful in future studies of NMR thermal state entanglability. 


\section{Conclusion} \label{sec:conclusion}

In summary, we have obtained a bound on the entanglability of thermal states in liquid-state NMR by classifying the entanglement of Bell-transformed thermal states.  These results are confirmed by more direct calculations of NPT bounds for two specific bipartite splits.

This paper sought to investigate two related issues: 1) the nature of mixed state entanglement in quantum computation and 2) how to most efficiently achieve entanglement in a liquid-state NMR system.  We have made some progress on both accounts.  Our tightest lower bound on thermal states scales as $N \sim 1/\alpha$ and provides us with a sense of the resource tradeoffs that must be made to achieve entanglement.  Moreover, this result explicitly shows that thermal states are more easily entangled than effective pure states.  Our bound identifies entanglable thermal states in $N$-$\alpha$ space outside the best bound on entanglable effective pure states.  Assuming the largest reasonable initial polarization ($\alpha = 4 \times 10^{-5}$), we find that a thermal state must have $N \leq 22,305$ to be entanglable, a limit that improves upon the D\"ur-Cirac constraint of $N \leq 50,000$ for effective pure states.  

Our methods exploit the symmetry of Bell unitaries, but there is considerable room for improvement, especially if these studies of liquid-state NMR entanglement are to motivate practical experiments.  We have suggested a majorization-based approach that uses a state $\rho^\prime$ of form $\rho_N$ to bound the entanglability of thermal state without having to choose a specific unitary.  The missing piece is an efficient construction of $\rho^\prime$ --- a task we leave for future work.



\begin{acknowledgments}
We express gratitude to Julia Kempe for bringing Ref.~\cite{Dur00a} to our attention and thank Andrew Cross and Joshua Powell for their help in the preliminary numerical calculations.  T.~M.~Y. acknowledges support from the Department of Defense through a NDSEG fellowship.  This work was partially funded by the NSF Center for Bits and Atoms under contract CCR-0122419.
\end{acknowledgments}


\begin{thebibliography}{52}
\expandafter\ifx\csname natexlab\endcsname\relax\def\natexlab#1{#1}\fi
\expandafter\ifx\csname bibnamefont\endcsname\relax
  \def\bibnamefont#1{#1}\fi
\expandafter\ifx\csname bibfnamefont\endcsname\relax
  \def\bibfnamefont#1{#1}\fi
\expandafter\ifx\csname citenamefont\endcsname\relax
  \def\citenamefont#1{#1}\fi
\expandafter\ifx\csname url\endcsname\relax
  \def\url#1{\texttt{#1}}\fi
\expandafter\ifx\csname urlprefix\endcsname\relax\def\urlprefix{URL }\fi
\providecommand{\bibinfo}[2]{#2}
\providecommand{\eprint}[2][]{\url{#2}}

\bibitem[{\citenamefont{Bennett et~al.}(1993)\citenamefont{Bennett, Brassard,
  Cr\'{e}peau, Jozsa, Peres, and Wootters}}]{Bennett93a}
\bibinfo{author}{\bibfnamefont{C.~H.} \bibnamefont{Bennett}},
  \bibinfo{author}{\bibfnamefont{G.}~\bibnamefont{Brassard}},
  \bibinfo{author}{\bibfnamefont{C.}~\bibnamefont{Cr\'{e}peau}},
  \bibinfo{author}{\bibfnamefont{R.}~\bibnamefont{Jozsa}},
  \bibinfo{author}{\bibfnamefont{A.}~\bibnamefont{Peres}}, \bibnamefont{and}
  \bibinfo{author}{\bibfnamefont{W.~K.} \bibnamefont{Wootters}},
  \bibinfo{journal}{Phys.~Rev.~Lett.} \textbf{\bibinfo{volume}{70}},
  \bibinfo{pages}{1895} (\bibinfo{year}{1993}).

\bibitem[{\citenamefont{Bennett
  et~al.}(1996{\natexlab{a}})\citenamefont{Bennett, DiVincenzo, Smolin, and
  Wootters}}]{Bennett96a}
\bibinfo{author}{\bibfnamefont{C.~H.} \bibnamefont{Bennett}},
  \bibinfo{author}{\bibfnamefont{D.~P.} \bibnamefont{DiVincenzo}},
  \bibinfo{author}{\bibfnamefont{J.~A.} \bibnamefont{Smolin}},
  \bibnamefont{and} \bibinfo{author}{\bibfnamefont{W.~K.}
  \bibnamefont{Wootters}}, \bibinfo{journal}{Phys.~Rev.~A.}
  \textbf{\bibinfo{volume}{54}}, \bibinfo{pages}{3824}
  (\bibinfo{year}{1996}{\natexlab{a}}).

\bibitem[{\citenamefont{Bennett
  et~al.}(1996{\natexlab{b}})\citenamefont{Bennett, Brassard, Popescu,
  Schumacher, Smolin, and Wootters}}]{Bennett96b}
\bibinfo{author}{\bibfnamefont{C.~H.} \bibnamefont{Bennett}},
  \bibinfo{author}{\bibfnamefont{G.}~\bibnamefont{Brassard}},
  \bibinfo{author}{\bibfnamefont{S.}~\bibnamefont{Popescu}},
  \bibinfo{author}{\bibfnamefont{B.}~\bibnamefont{Schumacher}},
  \bibinfo{author}{\bibfnamefont{J.~A.} \bibnamefont{Smolin}},
  \bibnamefont{and} \bibinfo{author}{\bibfnamefont{W.~K.}
  \bibnamefont{Wootters}}, \bibinfo{journal}{Phys.~Rev.~Lett.}
  \textbf{\bibinfo{volume}{76}}, \bibinfo{pages}{722}
  (\bibinfo{year}{1996}{\natexlab{b}}).

\bibitem[{\citenamefont{Azuma et~al.}(2001)\citenamefont{Azuma, Bose, and
  Vedral}}]{Azuma01a}
\bibinfo{author}{\bibfnamefont{H.}~\bibnamefont{Azuma}},
  \bibinfo{author}{\bibfnamefont{S.}~\bibnamefont{Bose}}, \bibnamefont{and}
  \bibinfo{author}{\bibfnamefont{V.}~\bibnamefont{Vedral}},
  \bibinfo{journal}{Phys.~Rev.~A.} \textbf{\bibinfo{volume}{64}},
  \bibinfo{pages}{062308} (\bibinfo{year}{2001}).

\bibitem[{\citenamefont{Braunstein and Pati}(2002)}]{Braunstein02a}
\bibinfo{author}{\bibfnamefont{S.~L.} \bibnamefont{Braunstein}}
  \bibnamefont{and} \bibinfo{author}{\bibfnamefont{A.~K.} \bibnamefont{Pati}},
  \bibinfo{journal}{Quantum Information and Computation}
  \textbf{\bibinfo{volume}{2}}, \bibinfo{pages}{399} (\bibinfo{year}{2002}).

\bibitem[{\citenamefont{Ekert and Jozsa}(1998)}]{Ekert98a}
\bibinfo{author}{\bibfnamefont{A.}~\bibnamefont{Ekert}} \bibnamefont{and}
  \bibinfo{author}{\bibfnamefont{R.}~\bibnamefont{Jozsa}},
  \bibinfo{journal}{Proc.~R.~Soc.~London, Ser.~A}
  \textbf{\bibinfo{volume}{356}}, \bibinfo{pages}{1769} (\bibinfo{year}{1998}).

\bibitem[{\citenamefont{Linden and Popescu}(2001)}]{Linden01a}
\bibinfo{author}{\bibfnamefont{N.}~\bibnamefont{Linden}} \bibnamefont{and}
  \bibinfo{author}{\bibfnamefont{S.}~\bibnamefont{Popescu}},
  \bibinfo{journal}{Phys.~Rev.~Lett.} \textbf{\bibinfo{volume}{87}},
  \bibinfo{pages}{047901} (\bibinfo{year}{2001}).

\bibitem[{\citenamefont{Jozsa and Linden}(2003)}]{Jozsa03a}
\bibinfo{author}{\bibfnamefont{R.}~\bibnamefont{Jozsa}} \bibnamefont{and}
  \bibinfo{author}{\bibfnamefont{N.}~\bibnamefont{Linden}},
  \bibinfo{journal}{Proc.~R.~Soc.~London, Ser.~A}
  \textbf{\bibinfo{volume}{459}}, \bibinfo{pages}{1471} (\bibinfo{year}{2003}).

\bibitem[{\citenamefont{Jones and Mosca}(1998)}]{Jones98a}
\bibinfo{author}{\bibfnamefont{J.~A.} \bibnamefont{Jones}} \bibnamefont{and}
  \bibinfo{author}{\bibfnamefont{M.}~\bibnamefont{Mosca}},
  \bibinfo{journal}{J.~Chem.~Phys.} \textbf{\bibinfo{volume}{109}},
  \bibinfo{pages}{1648} (\bibinfo{year}{1998}).

\bibitem[{\citenamefont{Jones et~al.}(1998)\citenamefont{Jones, Mosca, and
  Hansen}}]{Jones98b}
\bibinfo{author}{\bibfnamefont{J.~A.} \bibnamefont{Jones}},
  \bibinfo{author}{\bibfnamefont{M.}~\bibnamefont{Mosca}}, \bibnamefont{and}
  \bibinfo{author}{\bibfnamefont{R.~H.} \bibnamefont{Hansen}},
  \bibinfo{journal}{Nature (London)} \textbf{\bibinfo{volume}{393}},
  \bibinfo{pages}{344} (\bibinfo{year}{1998}).

\bibitem[{\citenamefont{Chuang et~al.}(1998{\natexlab{a}})\citenamefont{Chuang,
  Vandersypen, Zhou, Leung, and Lloyd}}]{Chuang98a}
\bibinfo{author}{\bibfnamefont{I.~L.} \bibnamefont{Chuang}},
  \bibinfo{author}{\bibfnamefont{L.~M.~K.} \bibnamefont{Vandersypen}},
  \bibinfo{author}{\bibfnamefont{X.}~\bibnamefont{Zhou}},
  \bibinfo{author}{\bibfnamefont{D.~W.} \bibnamefont{Leung}}, \bibnamefont{and}
  \bibinfo{author}{\bibfnamefont{S.}~\bibnamefont{Lloyd}},
  \bibinfo{journal}{Nature (London)} \textbf{\bibinfo{volume}{393}},
  \bibinfo{pages}{143} (\bibinfo{year}{1998}{\natexlab{a}}).

\bibitem[{\citenamefont{Chuang et~al.}(1998{\natexlab{b}})\citenamefont{Chuang,
  Gershenfeld, and Kubinec}}]{Chuang98b}
\bibinfo{author}{\bibfnamefont{I.~L.} \bibnamefont{Chuang}},
  \bibinfo{author}{\bibfnamefont{N.}~\bibnamefont{Gershenfeld}},
  \bibnamefont{and} \bibinfo{author}{\bibfnamefont{M.}~\bibnamefont{Kubinec}},
  \bibinfo{journal}{Phys.~Rev.~Lett.} \textbf{\bibinfo{volume}{80}},
  \bibinfo{pages}{3408} (\bibinfo{year}{1998}{\natexlab{b}}).

\bibitem[{\citenamefont{Vandersypen et~al.}(2001)\citenamefont{Vandersypen,
  Steffen, Breyta, Yannoni, Sherwood, and Chuang}}]{Vandersypen01a}
\bibinfo{author}{\bibfnamefont{L.~M.~K.} \bibnamefont{Vandersypen}},
  \bibinfo{author}{\bibfnamefont{M.}~\bibnamefont{Steffen}},
  \bibinfo{author}{\bibfnamefont{G.}~\bibnamefont{Breyta}},
  \bibinfo{author}{\bibfnamefont{C.~S.} \bibnamefont{Yannoni}},
  \bibinfo{author}{\bibfnamefont{M.}~\bibnamefont{Sherwood}}, \bibnamefont{and}
  \bibinfo{author}{\bibfnamefont{I.~L.} \bibnamefont{Chuang}},
  \bibinfo{journal}{Nature (London)} \textbf{\bibinfo{volume}{414}},
  \bibinfo{pages}{883} (\bibinfo{year}{2001}).

\bibitem[{\citenamefont{Braunstein et~al.}(1999)\citenamefont{Braunstein,
  Caves, Jozsa, Linden, Popescu, and Schack}}]{Braunstein99a}
\bibinfo{author}{\bibfnamefont{S.~L.} \bibnamefont{Braunstein}},
  \bibinfo{author}{\bibfnamefont{C.~M.} \bibnamefont{Caves}},
  \bibinfo{author}{\bibfnamefont{R.}~\bibnamefont{Jozsa}},
  \bibinfo{author}{\bibfnamefont{N.}~\bibnamefont{Linden}},
  \bibinfo{author}{\bibfnamefont{S.}~\bibnamefont{Popescu}}, \bibnamefont{and}
  \bibinfo{author}{\bibfnamefont{R.}~\bibnamefont{Schack}},
  \bibinfo{journal}{Phys.~Rev.~Lett.} \textbf{\bibinfo{volume}{83}},
  \bibinfo{pages}{1054} (\bibinfo{year}{1999}).

\bibitem[{\citenamefont{Laflamme et~al.}(2002)\citenamefont{Laflamme, Cory,
  Nevgrevergne, and Viola}}]{Laflamme02a}
\bibinfo{author}{\bibfnamefont{R.}~\bibnamefont{Laflamme}},
  \bibinfo{author}{\bibfnamefont{D.~G.} \bibnamefont{Cory}},
  \bibinfo{author}{\bibfnamefont{C.}~\bibnamefont{Nevgrevergne}},
  \bibnamefont{and} \bibinfo{author}{\bibfnamefont{L.}~\bibnamefont{Viola}},
  \bibinfo{journal}{Quantum Information and Computation}
  \textbf{\bibinfo{volume}{2}}, \bibinfo{pages}{166} (\bibinfo{year}{2002}).

\bibitem[{\citenamefont{Meyer}(2000)}]{Meyer00a}
\bibinfo{author}{\bibfnamefont{D.~A.} \bibnamefont{Meyer}},
  \bibinfo{journal}{Phys.~Rev.~Lett.} \textbf{\bibinfo{volume}{85}},
  \bibinfo{pages}{2014} (\bibinfo{year}{2000}).

\bibitem[{\citenamefont{Knill et~al.}(1998)\citenamefont{Knill, Chuang, and
  Laflamme}}]{Knill98a}
\bibinfo{author}{\bibfnamefont{E.}~\bibnamefont{Knill}},
  \bibinfo{author}{\bibfnamefont{I.}~\bibnamefont{Chuang}}, \bibnamefont{and}
  \bibinfo{author}{\bibfnamefont{R.}~\bibnamefont{Laflamme}},
  \bibinfo{journal}{Phys.~Rev.~A.} \textbf{\bibinfo{volume}{57}},
  \bibinfo{pages}{3348} (\bibinfo{year}{1998}).

\bibitem[{\citenamefont{Schack and Caves}(1999)}]{Schack99a}
\bibinfo{author}{\bibfnamefont{R.}~\bibnamefont{Schack}} \bibnamefont{and}
  \bibinfo{author}{\bibfnamefont{C.~M.} \bibnamefont{Caves}},
  \bibinfo{journal}{Phys.~Rev.~A.} \textbf{\bibinfo{volume}{60}},
  \bibinfo{pages}{4354} (\bibinfo{year}{1999}).

\bibitem[{\citenamefont{Menicucci and Caves}(2002)}]{Menicucci02a}
\bibinfo{author}{\bibfnamefont{N.~C.} \bibnamefont{Menicucci}}
  \bibnamefont{and} \bibinfo{author}{\bibfnamefont{C.~M.} \bibnamefont{Caves}},
  \bibinfo{journal}{Phys.~Rev.~Lett.} \textbf{\bibinfo{volume}{88}},
  \bibinfo{pages}{167901} (\bibinfo{year}{2002}).

\bibitem[{\citenamefont{Nielsen}(1998)}]{Nielsen98a}
\bibinfo{author}{\bibfnamefont{M.~A.} \bibnamefont{Nielsen}}, Ph.D. thesis,
  \bibinfo{school}{University of New Mexico} (\bibinfo{year}{1998}).

\bibitem[{\citenamefont{Anwar et~al.}(2004{\natexlab{a}})\citenamefont{Anwar,
  Jones, Blazina, Duckett, and Carteret}}]{Anwar04b}
\bibinfo{author}{\bibfnamefont{M.~S.} \bibnamefont{Anwar}},
  \bibinfo{author}{\bibfnamefont{J.~A.} \bibnamefont{Jones}},
  \bibinfo{author}{\bibfnamefont{D.}~\bibnamefont{Blazina}},
  \bibinfo{author}{\bibfnamefont{S.~B.} \bibnamefont{Duckett}},
  \bibnamefont{and} \bibinfo{author}{\bibfnamefont{H.~A.}
  \bibnamefont{Carteret}}, \bibinfo{journal}{eprint quant-ph/0406044}
  (\bibinfo{year}{2004}{\natexlab{a}}).

\bibitem[{\citenamefont{Anwar et~al.}(2004{\natexlab{b}})\citenamefont{Anwar,
  Blazina, Carteret, Duckett, and Jones}}]{Anwar04c}
\bibinfo{author}{\bibfnamefont{M.~S.} \bibnamefont{Anwar}},
  \bibinfo{author}{\bibfnamefont{D.}~\bibnamefont{Blazina}},
  \bibinfo{author}{\bibfnamefont{H.~A.} \bibnamefont{Carteret}},
  \bibinfo{author}{\bibfnamefont{S.~B.} \bibnamefont{Duckett}},
  \bibnamefont{and} \bibinfo{author}{\bibfnamefont{J.~A.} \bibnamefont{Jones}},
  \bibinfo{journal}{eprint quant-ph/0407091}
  (\bibinfo{year}{2004}{\natexlab{b}}).

\bibitem[{\citenamefont{Anwar et~al.}(2004{\natexlab{c}})\citenamefont{Anwar,
  Blazina, Carteret, Duckett, Halstead, Jones, Kozak, and Taylor}}]{Anwar04a}
\bibinfo{author}{\bibfnamefont{M.~S.} \bibnamefont{Anwar}},
  \bibinfo{author}{\bibfnamefont{D.}~\bibnamefont{Blazina}},
  \bibinfo{author}{\bibfnamefont{H.~A.} \bibnamefont{Carteret}},
  \bibinfo{author}{\bibfnamefont{S.~B.} \bibnamefont{Duckett}},
  \bibinfo{author}{\bibfnamefont{T.~K.} \bibnamefont{Halstead}},
  \bibinfo{author}{\bibfnamefont{J.~A.} \bibnamefont{Jones}},
  \bibinfo{author}{\bibfnamefont{C.~M.} \bibnamefont{Kozak}}, \bibnamefont{and}
  \bibinfo{author}{\bibfnamefont{R.~J.~K.} \bibnamefont{Taylor}},
  \bibinfo{journal}{Phys.~Rev.~Lett.} \textbf{\bibinfo{volume}{93}},
  \bibinfo{pages}{040501} (\bibinfo{year}{2004}{\natexlab{c}}).

\bibitem[{\citenamefont{D\"ur and Cirac}(2000)}]{Dur00a}
\bibinfo{author}{\bibfnamefont{W.}~\bibnamefont{D\"ur}} \bibnamefont{and}
  \bibinfo{author}{\bibfnamefont{J.~I.} \bibnamefont{Cirac}},
  \bibinfo{journal}{Phys.~Rev.~A.} \textbf{\bibinfo{volume}{61}},
  \bibinfo{pages}{042314} (\bibinfo{year}{2000}).

\bibitem[{\citenamefont{Peres}(1996)}]{Peres96a}
\bibinfo{author}{\bibfnamefont{A.}~\bibnamefont{Peres}},
  \bibinfo{journal}{Phys.~Rev.~Lett.} \textbf{\bibinfo{volume}{77}},
  \bibinfo{pages}{1413} (\bibinfo{year}{1996}).

\bibitem[{\citenamefont{Horodecki et~al.}(1996)\citenamefont{Horodecki,
  Horodecki, and Horodecki}}]{Horodecki96f}
\bibinfo{author}{\bibfnamefont{M.}~\bibnamefont{Horodecki}},
  \bibinfo{author}{\bibfnamefont{P.}~\bibnamefont{Horodecki}},
  \bibnamefont{and}
  \bibinfo{author}{\bibfnamefont{R.}~\bibnamefont{Horodecki}},
  \bibinfo{journal}{Phys.~Lett.~A} \textbf{\bibinfo{volume}{223}},
  \bibinfo{pages}{1} (\bibinfo{year}{1996}).

\bibitem[{\citenamefont{Gershenfeld and Chuang}(1997)}]{Gershenfeld97a}
\bibinfo{author}{\bibfnamefont{N.}~\bibnamefont{Gershenfeld}} \bibnamefont{and}
  \bibinfo{author}{\bibfnamefont{I.~L.} \bibnamefont{Chuang}},
  \bibinfo{journal}{Science} \textbf{\bibinfo{volume}{275}},
  \bibinfo{pages}{350} (\bibinfo{year}{1997}).

\bibitem[{\citenamefont{Cory et~al.}(1997)\citenamefont{Cory, Fahmy, and
  Havel}}]{Cory97a}
\bibinfo{author}{\bibfnamefont{D.~G.} \bibnamefont{Cory}},
  \bibinfo{author}{\bibfnamefont{A.~F.} \bibnamefont{Fahmy}}, \bibnamefont{and}
  \bibinfo{author}{\bibfnamefont{T.~F.} \bibnamefont{Havel}},
  \bibinfo{journal}{Proc.~Natl.~Acad.~Sci.~U.S.A.}
  \textbf{\bibinfo{volume}{94}}, \bibinfo{pages}{1634} (\bibinfo{year}{1997}).

\bibitem[{\citenamefont{Gurvits and Barnum}(2004)}]{Gurvits04a}
\bibinfo{author}{\bibfnamefont{L.}~\bibnamefont{Gurvits}} \bibnamefont{and}
  \bibinfo{author}{\bibfnamefont{H.}~\bibnamefont{Barnum}},
  \bibinfo{journal}{eprint quant-ph/0409095}  (\bibinfo{year}{2004}).

\bibitem[{\citenamefont{Vidal and Tarrach}(1999)}]{Vidal99a}
\bibinfo{author}{\bibfnamefont{G.}~\bibnamefont{Vidal}} \bibnamefont{and}
  \bibinfo{author}{\bibfnamefont{R.}~\bibnamefont{Tarrach}},
  \bibinfo{journal}{Phys.~Rev.~A.} \textbf{\bibinfo{volume}{59}},
  \bibinfo{pages}{141} (\bibinfo{year}{1999}).

\bibitem[{\citenamefont{$\dot{Z}$yczkowski
  et~al.}(1998)\citenamefont{$\dot{Z}$yczkowski, Horodecki, Sanpera, and
  Lewenstein}}]{Zyczkowski98a}
\bibinfo{author}{\bibfnamefont{K.}~\bibnamefont{$\dot{Z}$yczkowski}},
  \bibinfo{author}{\bibfnamefont{P.}~\bibnamefont{Horodecki}},
  \bibinfo{author}{\bibfnamefont{A.}~\bibnamefont{Sanpera}}, \bibnamefont{and}
  \bibinfo{author}{\bibfnamefont{M.}~\bibnamefont{Lewenstein}},
  \bibinfo{journal}{Phys.~Rev.~A.} \textbf{\bibinfo{volume}{58}},
  \bibinfo{pages}{883} (\bibinfo{year}{1998}).

\bibitem[{\citenamefont{Yamaguchi and Yamamoto}(1999)}]{Yamaguchi99a}
\bibinfo{author}{\bibfnamefont{F.}~\bibnamefont{Yamaguchi}} \bibnamefont{and}
  \bibinfo{author}{\bibfnamefont{Y.}~\bibnamefont{Yamamoto}},
  \bibinfo{journal}{Appl.~Phys.~A} \textbf{\bibinfo{volume}{68}},
  \bibinfo{pages}{1} (\bibinfo{year}{1999}).

\bibitem[{\citenamefont{Cory et~al.}(2000)\citenamefont{Cory, Laflamme, Knill,
  Viola, Havel, Boulant, Boutis, Fortunato, Lloyd, Martinez et~al.}}]{Cory00a}
\bibinfo{author}{\bibfnamefont{D.~G.} \bibnamefont{Cory}},
  \bibinfo{author}{\bibfnamefont{R.}~\bibnamefont{Laflamme}},
  \bibinfo{author}{\bibfnamefont{E.}~\bibnamefont{Knill}},
  \bibinfo{author}{\bibfnamefont{L.}~\bibnamefont{Viola}},
  \bibinfo{author}{\bibfnamefont{T.~F.} \bibnamefont{Havel}},
  \bibinfo{author}{\bibfnamefont{N.}~\bibnamefont{Boulant}},
  \bibinfo{author}{\bibfnamefont{G.}~\bibnamefont{Boutis}},
  \bibinfo{author}{\bibfnamefont{E.}~\bibnamefont{Fortunato}},
  \bibinfo{author}{\bibfnamefont{S.}~\bibnamefont{Lloyd}},
  \bibinfo{author}{\bibfnamefont{R.}~\bibnamefont{Martinez}},
  \bibnamefont{et~al.}, \bibinfo{journal}{Fortschr.~Phys.}
  \textbf{\bibinfo{volume}{48}}, \bibinfo{pages}{875} (\bibinfo{year}{2000}).

\bibitem[{\citenamefont{Verhulst et~al.}(2001)\citenamefont{Verhulst, Liivak,
  Sherwood, Vieth, and Chuang}}]{Verhulst01a}
\bibinfo{author}{\bibfnamefont{A.~S.} \bibnamefont{Verhulst}},
  \bibinfo{author}{\bibfnamefont{O.}~\bibnamefont{Liivak}},
  \bibinfo{author}{\bibfnamefont{M.~H.} \bibnamefont{Sherwood}},
  \bibinfo{author}{\bibfnamefont{H.-M.} \bibnamefont{Vieth}}, \bibnamefont{and}
  \bibinfo{author}{\bibfnamefont{I.~L.} \bibnamefont{Chuang}},
  \bibinfo{journal}{Appl.~Phys.~Lett.} \textbf{\bibinfo{volume}{79}},
  \bibinfo{pages}{2480} (\bibinfo{year}{2001}).

\bibitem[{\citenamefont{Zook et~al.}(2002)\citenamefont{Zook, Adhyaru, and
  Bowers}}]{Zook02a}
\bibinfo{author}{\bibfnamefont{A.~L.} \bibnamefont{Zook}},
  \bibinfo{author}{\bibfnamefont{B.~B.} \bibnamefont{Adhyaru}},
  \bibnamefont{and} \bibinfo{author}{\bibfnamefont{C.~R.}
  \bibnamefont{Bowers}}, \bibinfo{journal}{J.~Magn.~Reson.}
  \textbf{\bibinfo{volume}{159}}, \bibinfo{pages}{172} (\bibinfo{year}{2002}).

\bibitem[{\citenamefont{Jones}(2000)}]{Jones00a}
\bibinfo{author}{\bibfnamefont{J.~A.} \bibnamefont{Jones}},
  \bibinfo{journal}{Fortsch.~Phys.} \textbf{\bibinfo{volume}{48}},
  \bibinfo{pages}{909} (\bibinfo{year}{2000}).

\bibitem[{\citenamefont{Linden et~al.}(1999)\citenamefont{Linden, Kup\v{c}e,
  and Freeman}}]{Linden99a}
\bibinfo{author}{\bibfnamefont{N.}~\bibnamefont{Linden}},
  \bibinfo{author}{\bibfnamefont{E.}~\bibnamefont{Kup\v{c}e}},
  \bibnamefont{and} \bibinfo{author}{\bibfnamefont{R.}~\bibnamefont{Freeman}},
  \bibinfo{journal}{Chem.~Phys.~Lett.} \textbf{\bibinfo{volume}{311}},
  \bibinfo{pages}{321} (\bibinfo{year}{1999}).

\bibitem[{\citenamefont{Lloyd}(1993)}]{Lloyd93a}
\bibinfo{author}{\bibfnamefont{S.}~\bibnamefont{Lloyd}},
  \bibinfo{journal}{Science} \textbf{\bibinfo{volume}{261}},
  \bibinfo{pages}{1569} (\bibinfo{year}{1993}).

\bibitem[{\citenamefont{Sinha et~al.}(2001)\citenamefont{Sinha, Mahesh,
  Ramanathan, and Kumar}}]{Sinha01a}
\bibinfo{author}{\bibfnamefont{N.}~\bibnamefont{Sinha}},
  \bibinfo{author}{\bibfnamefont{T.~S.} \bibnamefont{Mahesh}},
  \bibinfo{author}{\bibfnamefont{K.}~\bibnamefont{Ramanathan}},
  \bibnamefont{and} \bibinfo{author}{\bibfnamefont{A.}~\bibnamefont{Kumar}},
  \bibinfo{journal}{J.~Chem.~Phys.} p. \bibinfo{pages}{4415}
  (\bibinfo{year}{2001}).

\bibitem[{\citenamefont{Murali et~al.}(2002)\citenamefont{Murali, Sinha,
  Mahesh, Levitt, Ramanathan, and Kumar}}]{Murali02a}
\bibinfo{author}{\bibfnamefont{K.~V.~R.~M.} \bibnamefont{Murali}},
  \bibinfo{author}{\bibfnamefont{N.}~\bibnamefont{Sinha}},
  \bibinfo{author}{\bibfnamefont{T.~S.} \bibnamefont{Mahesh}},
  \bibinfo{author}{\bibfnamefont{M.~H.} \bibnamefont{Levitt}},
  \bibinfo{author}{\bibfnamefont{K.~V.} \bibnamefont{Ramanathan}},
  \bibnamefont{and} \bibinfo{author}{\bibfnamefont{A.}~\bibnamefont{Kumar}},
  \bibinfo{journal}{Phys.~Rev.~A} \textbf{\bibinfo{volume}{66}},
  \bibinfo{pages}{022313} (\bibinfo{year}{2002}).

\bibitem[{\citenamefont{Schulman and Vazirani}(1999)}]{Schulman99a}
\bibinfo{author}{\bibfnamefont{L.~J.} \bibnamefont{Schulman}} \bibnamefont{and}
  \bibinfo{author}{\bibfnamefont{U.}~\bibnamefont{Vazirani}},
  \bibinfo{journal}{Proc. 31'st ACM STOC (Symp. Theory of Computing)} pp.
  \bibinfo{pages}{322--329} (\bibinfo{year}{1999}).

\bibitem[{\citenamefont{Boykin et~al.}(2002)\citenamefont{Boykin, Mor,
  Roychowdhury, Vatan, and Vrijen}}]{Boykin02a}
\bibinfo{author}{\bibfnamefont{P.~O.} \bibnamefont{Boykin}},
  \bibinfo{author}{\bibfnamefont{T.}~\bibnamefont{Mor}},
  \bibinfo{author}{\bibfnamefont{V.}~\bibnamefont{Roychowdhury}},
  \bibinfo{author}{\bibfnamefont{F.}~\bibnamefont{Vatan}}, \bibnamefont{and}
  \bibinfo{author}{\bibfnamefont{R.}~\bibnamefont{Vrijen}},
  \bibinfo{journal}{Proc.~Natl.~Acad.~Sci.~U.S.A.}
  \textbf{\bibinfo{volume}{99}}, \bibinfo{pages}{3388} (\bibinfo{year}{2002}).

\bibitem[{\citenamefont{Verstraete et~al.}(2001)\citenamefont{Verstraete,
  Audenaert, and {De Moor}}}]{Verstraete01a}
\bibinfo{author}{\bibfnamefont{F.}~\bibnamefont{Verstraete}},
  \bibinfo{author}{\bibfnamefont{K.}~\bibnamefont{Audenaert}},
  \bibnamefont{and} \bibinfo{author}{\bibfnamefont{B.}~\bibnamefont{{De
  Moor}}}, \bibinfo{journal}{Phys.~Rev.~A.} \textbf{\bibinfo{volume}{64}},
  \bibinfo{pages}{012316} (\bibinfo{year}{2001}).

\bibitem[{\citenamefont{Nielsen et~al.}(2003)\citenamefont{Nielsen, Dawson,
  Dodd, Gilchrist, Mortimer, Osborne, Bremner, Harrow, and Hines}}]{Nielsen03a}
\bibinfo{author}{\bibfnamefont{M.~A.} \bibnamefont{Nielsen}},
  \bibinfo{author}{\bibfnamefont{C.~M.} \bibnamefont{Dawson}},
  \bibinfo{author}{\bibfnamefont{J.~L.} \bibnamefont{Dodd}},
  \bibinfo{author}{\bibfnamefont{A.}~\bibnamefont{Gilchrist}},
  \bibinfo{author}{\bibfnamefont{D.}~\bibnamefont{Mortimer}},
  \bibinfo{author}{\bibfnamefont{T.~J.} \bibnamefont{Osborne}},
  \bibinfo{author}{\bibfnamefont{M.~J.} \bibnamefont{Bremner}},
  \bibinfo{author}{\bibfnamefont{A.~W.} \bibnamefont{Harrow}},
  \bibnamefont{and} \bibinfo{author}{\bibfnamefont{A.}~\bibnamefont{Hines}},
  \bibinfo{journal}{Phys.~Rev.~A.} \textbf{\bibinfo{volume}{67}},
  \bibinfo{pages}{052301} (\bibinfo{year}{2003}).

\bibitem[{\citenamefont{Vidal and Werner}(2002)}]{Vidal02a}
\bibinfo{author}{\bibfnamefont{G.}~\bibnamefont{Vidal}} \bibnamefont{and}
  \bibinfo{author}{\bibfnamefont{R.~F.} \bibnamefont{Werner}},
  \bibinfo{journal}{Phys.~Rev.~A.} \textbf{\bibinfo{volume}{65}},
  \bibinfo{pages}{032314} (\bibinfo{year}{2002}).

\bibitem[{\citenamefont{Vedral and Plenio}(1998)}]{Vedral98a}
\bibinfo{author}{\bibfnamefont{V.}~\bibnamefont{Vedral}} \bibnamefont{and}
  \bibinfo{author}{\bibfnamefont{M.~B.} \bibnamefont{Plenio}},
  \bibinfo{journal}{Phys.~Rev.~A.} \textbf{\bibinfo{volume}{57}},
  \bibinfo{pages}{1619} (\bibinfo{year}{1998}).

\bibitem[{\citenamefont{Nielsen}(1999)}]{Nielsen99a}
\bibinfo{author}{\bibfnamefont{M.~A.} \bibnamefont{Nielsen}},
  \bibinfo{journal}{Phys.~Rev.~Lett.} \textbf{\bibinfo{volume}{83}},
  \bibinfo{pages}{436} (\bibinfo{year}{1999}).

\bibitem[{\citenamefont{Nielsen and Vidal}(2001)}]{Nielsen02a}
\bibinfo{author}{\bibfnamefont{M.~A.} \bibnamefont{Nielsen}} \bibnamefont{and}
  \bibinfo{author}{\bibfnamefont{G.}~\bibnamefont{Vidal}},
  \bibinfo{journal}{Quantum Information and Computation}
  \textbf{\bibinfo{volume}{1}}, \bibinfo{pages}{76} (\bibinfo{year}{2001}).

\bibitem[{\citenamefont{Uhlmann}(1970)}]{Uhlmann70a}
\bibinfo{author}{\bibfnamefont{A.}~\bibnamefont{Uhlmann}},
  \bibinfo{journal}{Rep.~Math.~Phys.} \textbf{\bibinfo{volume}{1}},
  \bibinfo{pages}{147} (\bibinfo{year}{1970}).

\bibitem[{\citenamefont{Uhlmann}(1971)}]{Uhlmann71a}
\bibinfo{author}{\bibfnamefont{A.}~\bibnamefont{Uhlmann}},
  \bibinfo{journal}{Wiss.~Z.~Karl-Marx-Univ.~Leipzig}
  \textbf{\bibinfo{volume}{20}}, \bibinfo{pages}{633} (\bibinfo{year}{1971}).

\bibitem[{\citenamefont{Uhlmann}(1972)}]{Uhlmann72a}
\bibinfo{author}{\bibfnamefont{A.}~\bibnamefont{Uhlmann}},
  \bibinfo{journal}{Wiss.~Z.~Karl-Marx-Univ.~Leipzig}
  \textbf{\bibinfo{volume}{21}}, \bibinfo{pages}{421} (\bibinfo{year}{1972}).

\bibitem[{\citenamefont{Uhlmann}(1973)}]{Uhlmann73a}
\bibinfo{author}{\bibfnamefont{A.}~\bibnamefont{Uhlmann}},
  \bibinfo{journal}{Wiss.~Z.~Karl-Marx-Univ.~Leipzig}
  \textbf{\bibinfo{volume}{22}}, \bibinfo{pages}{139} (\bibinfo{year}{1973}).

\end{thebibliography}
\end{document}